\documentclass{ws-ijmpa}
\usepackage[super,compress]{cite}
\usepackage{hyperref}
\hypersetup{colorlinks,urlcolor=red,citecolor=red,linkcolor=red,filecolor=red}
\usepackage{breakurl}
\usepackage{url}
\usepackage{graphicx}
\usepackage{mathtools}
\usepackage{tabularx}
\usepackage{amsmath}
\usepackage{color}
\usepackage{amsfonts}
\usepackage{braket}
\usepackage{slashed}
\usepackage{xcolor}
\allowdisplaybreaks

\DeclareMathOperator{\tr}{tr}
\DeclareMathOperator{\Tr}{Tr}

\newcommand{\diag}{\text{diag}}
\newcommand{\instate}{|\text{in}\rangle}
\newcommand{\outstate}{|\text{out}\rangle}
\newcommand{\statein}{\langle\text{in}|}
\newcommand{\stateout}{\langle\text{out}|}
\newcommand{\llangle}{\langle\!\langle}
\newcommand{\rrangle}{\rangle\!\rangle}
\newcommand{\Sprop}{S_{\text{in}}^{\text{c}}}
\newcommand{\sprop}{S^{\text{c}}}

\newcommand{\calO}{\mathcal{O}}

\newcommand{\calF}{\mathcal{F}}

\newcommand{\D}{\mathcal{D}}
\newcommand{\appropto}{\mathrel{\vcenter{
  \offinterlineskip\halign{\hfil$##$\cr
    \propto\cr\noalign{\kern2pt}\sim\cr\noalign{\kern-2pt}}}}}
    
\begin{document}
\markboth{Patrick Copinger and Shi Pu}{Chirality Production 
with Mass Effects--Schwinger Pair Production and the Axial Ward Identity}

%
\catchline{}{}{}{}{}
%

\title{Chirality Production with Mass Effects--Schwinger 
Pair Production and the Axial Ward Identity}

\author{Patrick Copinger}

\address{Theory Center, Institute of Particle and Nuclear Studies,\\
High Energy Accelerator Research Organization (KEK)\\
1-1 Oho, Tsukuba, Ibaraki 305-0801, Japan\\ 
\vspace{0.5em}
Department of Physics and Center for Field Theory 
and Particle Physics,\\Fudan University, 220 Handan Rd.,
Shanghai 200433, China\\
copinger@post.kek.jp}

\author{Shi Pu}

\address{Department of Modern Physics and Interdisciplinary 
Center for Theoretical Study,\\University of Science and 
Technology of China, Hefei,
Anhui 230026, China\\
shipu@ustc.edu.cn}

\maketitle


\begin{abstract}

The anomalous generation of chirality with mass effects via
the axial Ward identity and its dependence on the Schwinger mechanism
is reviewed, utilizing parity violating homogeneous electromagnetic background
fields. The role vacuum asymptotic states play on 
the interpretation of expectation values is examined. It is discussed
that observables calculated with an in-out scattering matrix element
predict a scenario under Euclidean equilibrium. A notable ramification of
which is a vanishing of the chiral anomaly. In contrast, it is discussed
observables calculated under an in-in, or real-time, formalism predict
a scenario out-of equilibrium, and capture effects of mean produced 
particle anti-particle pairs due to the Schwinger mechanism. The out-of
equilibrium chiral anomaly is supplemented with exponential
quadratic mass suppression as anticipated for the Schwinger mechanism.
Similar behavior in and out-of equilibrium is reviewed for
applications including the chiral magnetic 
effect and chiral condensate.

\end{abstract}

\ccode{PACS numbers: 05.70.Ln, 11.30.Rd, 12.20.--m}


\section{Introduction}
\label{sec:introduction}

An anomaly manifests itself for systems with a symmetry that, while at
a classical level is realized, is actually broken at the quantum level.
For relativistic fermionic systems the chiral 
symmetry~\cite{BARRON1986423} gives rise to
such an anomaly, and the breaking of the chiral symmetry is of
paramount importance for several phenomena, notably including
imparting the bulk of the visible mass to 
the universe~\cite{Nambu:1961tp,Nambu:1961fr}. The direct
observation of the chiral anomaly, however, remains, and an 
essential application of the anomaly that may facilitate its
observation is the chiral magnetic effect (CME).

The CME is
 an electromagnetic current in the 
presence of and along the direction of
a magnetic field due to a net 
chirality~\cite{PhysRevD.78.074033}. Relativistic fermionic
dispersion relations are realizable in 2D and 3D condensed matter 
systems for Weyl and Dirac semimetals
\cite{PhysRevB.85.195320, PhysRevB.88.125427,Neupane2014,Liu2014}. 
And, in a Dirac semimetal the CME was thought to be 
observed~\cite{Li:2014bha}.
Even so,  it is still challenging to observe the CME 
in relativistic heavy-ion collisions due to huge background contributions, despite 
a strong magnetic field thought present--in fact, 
a field as large as $eB\sim m_{\pi}^2$ may be 
possible~\cite{Kharzeev:2013jha}--in off-central collisions. 

The presence of the other ingredient of the CME, namely a net chirality, 
in colliders comes with greater uncertainty. And it is an 
uncertainty one may mitigate with an improved 
understanding of how chirality is generated. We address the 
following three issues in this review:
\begin{enumerate}
\item Chirality imbalance frequently is inserted by hand, usually
by means of a chiral chemical potential. However, while a useful
theoretical tool, there are instances where such insertions are
inadequate. One such case lies with systems well out-of
equilibrium.
\item The behavior of the chiral anomaly and magnetic effect 
in and out-of equilibrium requires elucidation. The out-of
equilibrium case is prominent in heavy-ion collisions. There a 
glasma~\cite{Kharzeev:2001ev,Lappi:2006fp}, 
or dense gluonic state, is thought to give rise to
parity-violating flux tubes and the accompanying chiral
anomaly and CME~\cite{Fukushima:2010vw}.
\item Finally, what are the effects of a finite mass on the 
chiral anomaly and CME? While in high energy applications
it is common to dismiss the mass--e.g., a pseudoscalar term,
we will go on to argue that this dismissal is not subtle.
\end{enumerate}

The answer to the above questions can be addressed
through the Schwinger mechanism. In a background electric
field the quantum field theoretic (QFT) vacuum is thought to
be unstable against the creation of particle anti-particles
pairs through tunneling in what is known as the Schwinger 
mechanism. The QCD electric field analog is provided by 
chromo-electric flux tubes, whose breaking is facilitated 
through the Schwinger mechanism leading to 
hadronization~\cite{PhysRevD.20.179}. How might one 
furnish a net chirality from the Schwinger mechanism?
This is thought possible with a parallel strong magnetic field, 
setting up a parity-violating background. Then predicted
pairs of particles have their spins aligned with the magnetic 
field generating a net chirality~\cite{Fukushima:2010vw,Warringa:2012bq}.
This phenomenon has also been studied
numerically~\cite{Tanji:2010eu,Andres:2018ifx,Tanji:2018qws}.

The axial Ward identity provides an appropriate means of accessing
the chirality non-conservation for massive fermionic 
systems~\cite{PhysRev.177.2426,Bell:1969ts}, and is composed
of both a contribution due to quantum effects as well as a massive
pseudoscalar term. Taking expectation values of the axial Ward
identity, however, using standard treatments lead to
puzzling results in contrast to the picture of chirality generation
via the Schwinger mechanism; notably a conservation of chirality can be
found.

A clear identification of vacuum states and their expectation values 
provides a resolution~\cite{Copinger:2018ftr}. Usage of either 
in-out or in-in vacuum states predict decidedly different physical
scenarios. The expectation values of in-out vacuum states, 
used in standard approaches,
predict a scenario of Euclidean equilibrium.
However, the Schwinger mechanism--and hence chiral driven 
phenomena--is inherently out-of
equilibrium, and the real-time process of pair production
is not captured with an in-out formalism. It is, however, captured
utilizing an in-in, (or Schwinger-Keldysh), formalism. There
a chirality non-conservation is predicted out-of equilibrium 
in accordance with
the Schwinger mechanism, as evidenced by an exponential
quadratic mass suppression. This has implications for theories built
on an anomaly, e.g. for baryogenesis driven by a parity-violating 
inflaton~\cite{Domcke:2019qmm}, 
as well as experimental 
ramifications. In addition to the chiral anomaly by the axial Ward identity, 
we also examine the CME as well as the chiral
condensate.

Even though a system possesses no global net chirality, this does not mean
that it might not be present locally. This is thought to be the case in 
heavy-ion collisions~\cite{Kharzeev:2007jp,Bzdak:2011yy}. One way one
might characterize a local non-conservation of chirality is through
an examination of chiral density fluctuations, or a chiral susceptibility. 
Similar in and out-of equilibrium behavior, with characteristic Schwinger
mechanism signatures, is noted for such correlated
observables.

The enhancement of the chiral condensate, or rather a dynamically driven
mass, by a background magnetic field is known as magnetic 
catalysis~\cite{Klimenko:1992ch,Gusynin:1994re,Shovkovy:2012zn}. However,
how is the chiral condensate augmented by an electric field and the
Schwinger pair production process in and out-of equilibrium? We address
this here too. It is found the electric field serves to diminish the condensate,
and for strong enough fields may even negate the condensate all together. 

The structure of this review along with notations are given as follows:
To supplement the cursory look at the generation of chirality via the Schwinger
mechanism just offered, we give some background to the chiral anomaly and 
magnetic effects in Sec.~\ref{sec:cme_back} and to the Schwinger mechanism
in Sec.~\ref{sec:schwinger_back}. Then a heuristic picture of chirality
generation from the Schwinger mechanism is presented in 
Sec.~\ref{sec:heuristic}. Next we proceed with the identification
of vacuum states and their importance in the interpretation of expectation
values in Sec.~\ref{sec:vacuum}. Next, the application of vacuum states to
the generation of chirality through the axial Ward identity is discussed in 
Sec.~\ref{sec:awi}. Then extensions to other chiral anomaly related 
phenomena including the CME in Sec.~\ref{sec:cme}, 
and the chiral condensate in 
Sec.~\ref{sec:condensate} are discussed. Last, a conclusion to the review
is presented in Sec.~\ref{sec:conclusions}.

The following notations are used in this review: We use a mostly minus 
metric, $g=\diag(+,-,-,-)$, and whenever appropriate contracted
Lorentz indices are implicit, i.e., $A_\mu B^\mu =:AB$. Our 
covariant derivative reads $D_\mu = \partial_\mu + ieA_\mu$. And
we also use units such that $c=\hbar =1$. For gamma matrices
a Weyl representation is used:
\begin{equation}
   \gamma^{0}=\biggr(\begin{array}{cc} & I_{2}\\
   I_{2} \end{array}\biggl),
   \quad\gamma^{i}=\biggr(\begin{array}{cc} & \sigma^i\\
   -\sigma^i \end{array}\biggl),
   \quad\gamma_{5}=\biggr(\begin{array}{cc} -I_{2}\\ & I_{2}
   \end{array}\biggl)\,,
\label{eq:weyl_gamma}
\end{equation}
with $\sigma^i$ being the Pauli matrices. The spin tensor reads
$\sigma_{\mu\nu} = \frac{i}{2}[\gamma_\mu,\gamma_\nu]$.
Last, point split observables are evaluated with an averaged propagator
as in $S(x,x)= \tfrac{1}{2}\lim_{\epsilon\rightarrow0}
[S(x,x+\epsilon)+S(x+\epsilon,x)]$; elsewhere we take 
for Heaviside functions $\lim_{x\rightarrow 0}\theta(x) 
=[\theta(0^+)+\theta(0^-)]/2=1/2$.

\section{Chiral Anomaly and Magnetic Effect}
\label{sec:cme_back}

Anomalous phenomena are ubiquitous throughout physics and 
can be responsible for constraints on conserved currents, symmetries, and
spectrums of a theory. A clear manifestation of the anomaly in
experiment is provided by the decay of a neutral pion into two photons:
While classically forbidden, it was found at a quantum 
level~\cite{PhysRev.177.2426,Schwinger:1951nm} the process
be achievable.

One may characterize the anomaly of QCD with a topological 
$\theta$ term~\cite{Belavin197585}.
In QCD the Lagrangian can be supplemented with
$\frac{\theta}{32\pi}  \epsilon_{\mu\nu \alpha \beta} 
G^{\mu \nu a} G^{\alpha \beta a}$ for gluon field strength $G$
in $SU(3)$, and is both $\mathcal{P}$ (parity) 
and $\mathcal{CP}$ (charge-parity) violating. Unfortunately, 
there is no strong evidence to prove  such a term really exists in
 experiments--neutron dipole moments are
restricted to $|d_n| < 2.9 \times 10^{-26}$ $e$ cm~\cite{Baker:2006ts};
the theoretical underpinning of this problem is called the
``strong $\mathcal{CP}$ problem.'' However, there may be 
environments where topology may be present, if only locally. This is
thought to be the case in quark-gluon plasmas, giving rise to a 
axion-like topological term with space-time 
dependence\cite{Kharzeev:2009fn}, 
i.e., $\theta\rightarrow \theta(x)$. A consequence of which is a 
manifestation of an electromagnetic current, the CME.

\begin{figure}
	\centering
		\includegraphics[width=1.0\linewidth]{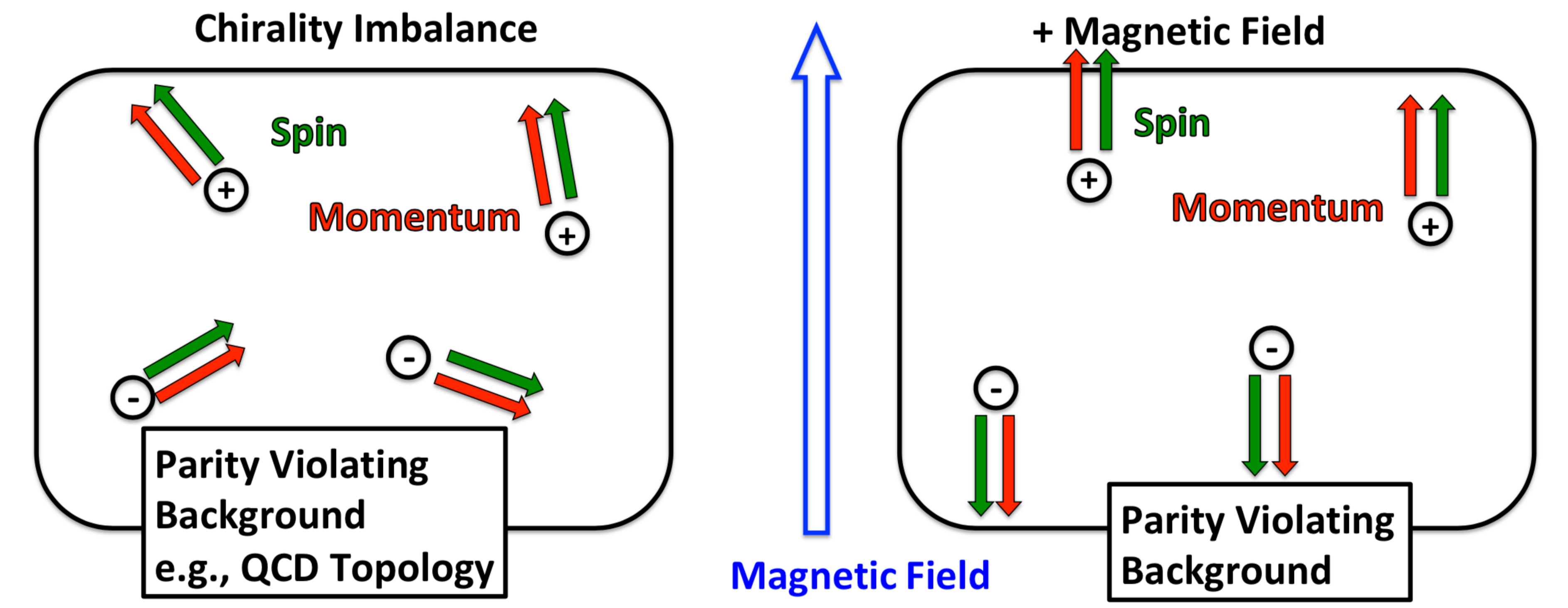}
		\caption{Diagram of the CME process. (Left) A non-conservation
		of chirality is dictated due to the chiral anomaly in a topologically
		non-trivial QCD background. Green are red arrows represent 
		spin and momentum vectors respectively. Plus and minus circles
		represent particles and anti-particles respectively. A net chirality
		difference of $\Delta N_5=4$ is shown. (Right) A strong magnetic
		field is then added, projecting the particles' spins and setting up an
		electromagnetic current in what is known as the CME.}
		\label{fig:cme}
\end{figure}

Due to a non-trivial topology, a chirality non-conservation is dictated
through the chiral anomaly. And in the context of a heavy-ion 
collision--for example--with a strong transverse to reaction 
plane magnetic field coupled with a net chirality, the CME is thought
to arise. Let us elaborate on the essential physics. See Fig.~\ref{fig:cme}
for the CME process. 

The left diagram depicts the chiral anomaly, 
where for a $\mathcal{P}$ violating background, such as for 
a topologically non-trivial background in a quark-gluon plasma, a net
chirality is furnished. We have made the assumption here of massless
particles, which entails that particle chirality and helicity be similar, 
and anti-particles have chirality opposite to their helicity. 
This amounts to a net chirality difference given as the total
number of particles plus antiparticles with right-handed helicity minus
the total with left-handed helicity~\cite{PhysRevD.78.074033}. 
Right (left)-handed helicity is given by parallel (anti-parallel) spin and 
momentum vectors, which we can denote figuratively as
$h_{R(L)}$ and $\bar{h}_{R(L)}$ for particles and antiparticles respectively.
 In other words $\Delta N_5 = \sum h_R + \bar{h}_R - h_L - \bar{h}_L$.
Then, in the right diagram in Fig.~\ref{fig:cme}, a strong magnetic
field is added. The lowest Landau level is occupied, projecting the 
particles' spins
to the direction of the magnetic field, giving rise to an electromagnetic
current, the CME. The process may also be understood as arising from the 
Dirac sea coupled with the chiral anomaly~\cite{Kharzeev:2009fn}, however
we reserve such discussions till Sec.~\ref{sec:heuristic}.

Enormous experimental effort has been carried out for the CME both in 
condensed matter and collider environments. Notably, the CME was thought
to be observed in a semimetal~\cite{Li:2014bha}. However, in heavy-ion
collision experiments, such as at the the large hadron collider (LHC) at 
CERN and 
the relativistic heavy ion collider (RHIC) at the Brookhaven National Laboratory,
the CME has yet to be confirmed.

Relativistic fermionic dispersion relations, as well as the chiral anomaly and 
CME, were thought producible in a number
of condensed matter environments including graphene and Weyl and 
Dirac semimetals, and spin-orbit coupled atomic 
gases~\cite{Huang:2015mga}. 
Let us describe and confine
our attention to the former. In contrast to a semi-conductor, the semimetal's/
graphene's valence and conduction bands possess a small overlap permitting
novel electronic and transport properties. Notably, a chirality can be governed
at the Weyl nodes, which serve as topological charges formed from a 
Berry's curvature in crystal quasi-momentum 
space~\cite{RevModPhys.82.1959,Chang_2008}. A relativistic massless
Weyl-like fermionic quasi-particle excitation and spectrum were discovered
in a 2+1 dimensional graphene~\cite{Novoselov666} and a 3+1 dimensional
semimetal~\cite{PhysRevB.85.195320,PhysRevB.88.125427,
PhysRevLett.113.027603,Liu864}. 
Accordingly, the anomaly was thought to be 
found in a Weyl semimetal~\cite{PhysRevX.5.031023}, as well as,
through a negative 
magnetoresistence~\cite{Son:2012bg} signature, the CME
was thought to be found in a semimetal~\cite{Li:2014bha}.

In the strong magnetic fields present in off-central heavy-ion collisions, 
the CME is
thought to be observable due to a local parity violation. 
Even though topological
fluctuations are not directly observable in collisions, charge asymmetries
of event-by-event correlations may be observable~\cite{KHARZEEV20161}.
Such measurements have been performed by groups STAR at the 
RHIC~\cite{PhysRevLett.103.251601} and ALICE at the 
LHC~\cite{PhysRevLett.110.012301}. However, while results are consistent
with local parity violation and the CME, background interference cannot be
mitigated, and thus verification of the CME in colliders cannot be accomplished
quite yet~\cite{KHARZEEV20161}.

In addition, the magnetic fields can also induce a chiral current, which is named 
Chiral Separation Effect (CSE) 
\cite{Kharzeev:2004ey,Kharzeev:2007jp,Fukushima:2008xe}. 
The collective modes of CME combined with CSE are
called Chiral Magnetic Waves, which are
also an important topic in relativistic heavy ion collisions 
\cite{Kharzeev:2010gd,Burnier:2011bf}. There are also many higher order 
nonlinear quantum phenomena related to electromagnetic fields, e.g., the chiral 
electric or Hall separation effects \cite{Huang:2013iia, Pu:2014cwa, 
Jiang:2014ura, Pu:2014fva} and other effects coupled to the gradient of 
temperature or chemical potentials 
\cite{Gorbar:2016qfh,Gorbar:2016sey,Chen:2016xtg,Ebihara:2017suq,
Hidaka:2018ekt,Bu:2019mow,Bu:2018psl,Bu:2018vcp,Liu:2020dxg}.

There are two ways to investigate the CME and other chiral transport 
phenomena. The microscopic description of the CME is called the chiral kinetic 
theory  (CKT), which is the quantum kinetic theory for the massless fermions. 
CKT can be derived from the path integrals 
\cite{Stephanov:2012ki,Chen:2013iga, Chen:2014cla,Chen:2015gta}, effective 
theories \cite{Son:2012zy, Manuel:2013zaa,Manuel:2014dza, Lin:2019ytz, 
Lin:2019fqo}, Wigner function approaches \cite{Chen:2012ca,Chen:2013dca, 
Hidaka:2016yjf,Hidaka:2017auj,Hidaka:2018mel,
Huang:2018wdl,Gao:2018wmr,Liu:2018xip,Yang:2020mtz} and world-line 
formalism \cite{Mueller:2017lzw, Mueller:2017arw}. Based on CKT, several 
numerical simulations for relativistic heavy ion collisions appear 
\cite{Sun:2016nig,Sun:2016mvh,Sun:2017xhx,Sun:2018idn,Sun:2018bjl,Zhou:2018rkh,Zhou:2019jag,Liu:2019krs}.
The quantum kinetic theory for the massive fermions 
\cite{Gao:2019znl,Wang:2019moi, Weickgenannt:2019dks,Li:2019qkf, 
Weickgenannt:2020sit,Sheng:2020oqs, Guo:2020zpa} and collisional terms 
\cite{Yang:2020hri,Weickgenannt:2020aaf} are also widely discussed recently. 

Another way to study CME is through the macroscopic effective theories based 
on the hydrodynamic equations coupled to the Maxwell's equations. One 
framework is named the relativistic magnetohydrodynamics 
\cite{Pu:2016ayh,Roy:2015kma,Pu:2016bxy,Pu:2016rdq,Roy:2017xtz, 
Siddique:2019gqh,Wang:2020qpx}, which has been widely used in 
astrophysics. The recent studies have been extended to the system in the 
presence of CME and chiral anomaly \cite{Siddique:2019gqh,Wang:2020qpx}. 
There are also several numerical simulations of ideal relativistic 
magnetohydrodynamics in the relativistic heavy ion collisions 
\cite{Inghirami:2016iru,Inghirami:2019mkc}. Beyond the ideal fluid, the second 
order magnetohydrodynamics including the dissipative effects are studied via 
the Grads momentum expansion \cite{Denicol:2018rbw,Denicol:2019iyh}.
Another macroscopic framework is named  Anomalous-Viscous Fluid Dynamics 
(AVFD) \cite{Jiang:2016wve,Shi:2017cpu,Shi:2017ucn}, where the magnetic 
fields are considered as the background fields. 
There are also many studies of the CME in a perturbation aspect of the 
quantum field theory
\cite{Feng:2017dom, Wu:2016dam, Lin:2018aon, Horvath:2019dvl, 
Feng:2018tpb} and the chiral charge fluctuation \cite{Hou:2017szz, Lin:2018nxj, 
Shi:2020uyb}. 
For more discussion on CME and other related topics, one can also see the 
recent reviews 
\cite{Kharzeev:2015znc,Liao:2014ava,Miransky:2015ava,Huang:2015oca, 
Skokov:2016yrj,Fukushima:2018grm,Bzdak:2019pkr,Zhao:2019hta, 
Wang:2018ygc,Liu:2020ymh,Gao:2020vbh}
and reference therein. Having explored some aspects of the anomaly and
the CME, let us explore the Schwinger mechanism, and then we
can establish their connection.

\section{Schwinger Mechanism}
\label{sec:schwinger_back}

In the presence of a strong background electric field, the QFT vacuum is
thought unstable against the production of particle anti-particles in what
is known as the Schwinger 
mechanism~\cite{sauter,Heisenberg:1935qt,Schwinger:1951nm}. The 
mechanism had its beginnings as a solution to the Klein 
paradox~\cite{Klein:1929zz}, which
highlighted the particle non-conserving properties of relativistic
QFTs. The Schwinger mechanism may be classified as a QFT 
instability; others include 
Hawking radiation~\cite{Hawking1975}, the Unruh 
effect~\cite{PhysRevD.14.870}, pair creation from inflation, e.g., in a 
Robertson Walker metric, and spontaneous symmetry 
breaking~\cite{Higgs:1964pj}. Moreover, the Schwinger mechanism 
may prove indispensable in that it might be used to access  such other 
QFT instabilities; e.g., 
gravitational effects in Hawking radiation could be 
mimicked~\cite{Dunne:2010zz}.
Schwinger pair production is thought to take place not only in QED but also in 
Yang-Mills theories~\cite{Yildiz:1979vv,Ambjorn1983340,Gyulassy:1986jq},
and is thought to lead to hadronization stemming from a breaking of 
chromoelectric flux tubes. 

\begin{figure}
\centering
  \includegraphics[width=0.4\columnwidth]{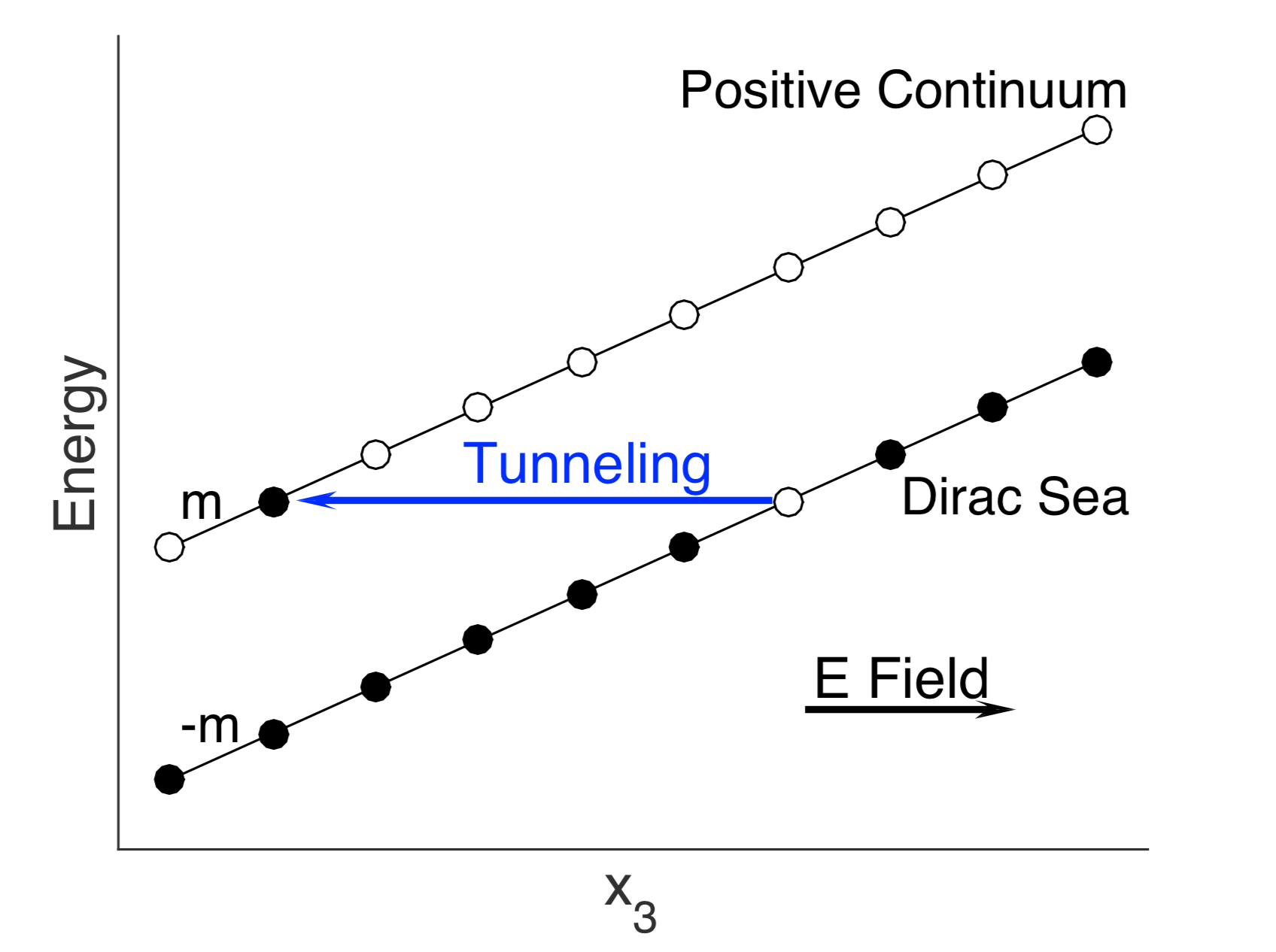}
  \caption{The Schwinger mechanism in a Dirac sea picture. Dirac spectrum
  is given for symbolic coordinate, $x_3$, the direction of electric field. Excitations
  (holes) represent particles (anti-particles) whose traversal of the mass gap 
  is made possible owing to the virtue of the electric field, that acts to tilt
  the spectrum.}
  \label{fig:sea_position}
\end{figure}

To understand the Schwinger pair production simply, let us make use of
a Dirac sea picture. See Fig.~\ref{fig:sea_position}. 
It can be seen that under an electric field, with strength $E$, that the Dirac
spectrum is tilted by $Ex_3$, allowing a quantum tunneling of the 
mass gap that creates a particle anti-particle pair. A characteristic tunneling
length must be passed that is proportional to the critical electric field. Not
only from a Dirac sea picture, but also from an intuitive classical perspective
can one understand Schwinger pair production. Consider a representation
of the vacuum as being composed of virtual particle anti-particle pairs in a 
condensate, then an electric field may impart work to the pair separating and 
accelerating them apart. 

One may identity a vacuum instability through inequivalent vacuum states
at asymptotic times, i.e., $\statein\neq\stateout$ at $t_{in}$ and $t_{out}$
respectively. An S matrix element calculation predicting the vacuum stay the 
vacuum provides a measure of a vacuum instability, or rather what is more is 
the calculation predicting anything but the vacuum appear in the out state: 
It is referred to as the vacuum non-persistence and is given by the probability
$\mathcal{P}\coloneqq1-|\langle \text{out}|\text{in}\rangle|^2$.
$\langle \text{out}|\text{in}\rangle$ here, confining our attention, 
is given by the QED
partition function in a background field,
\begin{equation}
   c_v\coloneqq\langle \text{out}|\text{in}\rangle
   =\int\mathcal{D}\bar{\psi}\mathcal{D}\psi
   \exp\bigl\{i\int d^{4}x[\bar{\psi}(i\slashed{D}-m)\psi]\bigr\}\,.
    \label{eq:cv}
\end{equation}
Then casting the partition function as an effective action, 
$e^{i\Gamma}\coloneqq c_v$, the vacuum non-persistence becomes
\begin{equation}
   \mathcal{P}\approx 2 \text{Im} \Gamma\,,
   \label{eq:probability}
\end{equation}
for small imaginary parts of $\Gamma$. Turning our attention to the 
Schwinger mechanism, the vacuum non-persistence in a homogeneous
electric field with strength, $E$, for fermions with mass, $m$, 
is~\cite{Schwinger:1951nm}
\begin{align}
   \mathcal{P}\appropto \exp\Bigl(\frac{-\pi m^2c^3}{eE\hbar}\Bigr)\,.
   \label{eq:critical_field}
\end{align}
Here SI units have been used to highlight the electric field strength 
required to see Schwinger pair production. Indeed the required field strength 
is large, in excess of modern capabilities, and therefore the Schwinger 
mechanism has yet to observed. But, special temporally inhomogeneous field
profiles have shown promise to overcome this difficulty.

Let us digress shortly on ``dynamically assisted'' 
fields~\cite{Schutzhold:2008pz}. Keldysh first found a temporal
inhomogeneity in the electric field would lessen the threshold for pair 
production~\cite{Keldysh_1964}, effectively relying on a combination
of pertubative and non-perturbative components. Then motivated by 
Keldysh's work a dynamically assisted combinatory field with both a 
strong amplitude and low 
frequency component as well as a weak amplitude and high frequency 
component
was studied, which dramatically improved the 
probability for pair 
production~\cite{Schutzhold:2008pz,PhysRevD.80.111301,
PhysRevD.81.025001,PhysRevD.81.085014}. The dynamical mechanism
and its spin-dependence have also been studied in a perturbative Furry 
picture~\cite{PhysRevD.99.056006,
PhysRevD.100.016013,PhysRevResearch.2.023257,
Torgrimsson:2017pzs,PhysRevD.99.096002}, 
as well as numerically~\cite{10.1093/ptep/ptz112}. A kinetic
theory with usage of the dynamical mechanism~\cite{Kohlfurst:2012rb}, and 
the momenta spectra~\cite{Orthaber:2011cm} were also
studied.

Eq.~\eqref{eq:critical_field} reflects an exponentially quadratic 
mass suppression,
and is the indicative feature of the Schwinger mechanism. In homogeneous
fields the factor should be present for observables where the 
Schwinger mechanism
plays a role. Also, let us use Eq.~\eqref{eq:critical_field} to illustrate the 
non-perturbative nature of the Schwinger mechanism; this can be seen from
the gauge coupling constant $e$. Schwinger pair production cannot be seen 
at any order in perturbation theory. Verification or falsification of the Schwinger
mechanism is highly sought and some environments thought capable include
condensed matter systems, high powered lasers, and heavy ion collisions.

Condensed matter environments such as for 
Weyl/Dirac semimetals~\cite{PhysRevB.92.085122}, 
semiconductors~\cite{PhysRevB.97.035203}, 
or graphene~\cite{PhysRevD.78.096009,PhysRevB.81.165431} 
for the potential observation
of the Schwinger mechanism are desirable due to a considerably 
reduced energy 
gap~\cite{PhysRevB.97.035203}. In condensed matter environments Schwinger
pair production is facilitated through a Landau Zener 
transition~\cite{10011873546,rspa.1932.0165}. In place of the 
positive continuum 
(Dirac sea) lies the conduction (valence) band. While semimetals and 
graphene have
little or no energy gap, a doped semimetal may possess a tunable 
gap~\cite{Kim723}.
Then for gap, $\Delta$, one can find for the condensed matter analog the 
non-persistence probability has the form
$\mathcal{P} \approx 
\exp(-\frac{\pi \Delta^2}{v_F \hbar eE})$~\cite{KANE1960181}. 
In addition to a
lessening of the critical field strength, inhomogeneous fields too may prove 
beneficial~\cite{PhysRevB.97.035203}, not only in condensed matter but also 
in QED.

Direct observation of the Schwinger mechanism is theoretically achievable in 
high powered lasers, and it is an essential task to pursue. Experimentation can 
be managed with either sole use of lasers or through a 
laser particle beam collision~\cite{PhysRevD.60.092004}. 
Strong QED is actively being studied at numerous high powered laser facilities 
including the Extreme Light Infrastructure (ELI) and the X-ray Free-Electron 
Laser Facility (XFEL) in DESY; experimental outlooks are provided in 
Refs.~\citen{Dunne:2010zz,Marklund:2008gj,Dunne:2008kc}. However,
peak electric fields produced are still order of magnitude, i.e. $\sim10^{-2}$,
lower than is the critical electric field required for Schwinger pair production, 
$\sim1.3\times10^{18}$ V/m~\cite{Ringwald:2003iv,Heinzl:2008an}. 
More so, realistic modeling of high powered laser beams are highly 
inhomogeneous
require numerical modeling; e.g., see Ref.~\citen{PhysRevE.92.023305}. 

On the other hand, the electromagnetic fields generated in relativistic 
heavy ion collisions are at the order of a few $m_\pi^2$ with $m_\pi$ the 
mass of the pion meson
\cite{Bloczynski:2012en,Deng:2012pc,Roy:2015coa,Li:2016tel}. 
Since  the quantum electromagnetic dynamics dominates in the 
ultra-peripheral collisions (UPC),
those experiments may provide a nice and possible platform to study the 
non-linear effects of QED
\cite{Adam:2019mby,Klein:2018fmp,Li:2019yzy,Zha:2018tlq,Zha:2018ywo}. 

With an understanding of the anomaly and Schwinger mechanism in hand,
let us examine more closely how chirality may be spawned through pair
production.

\section{Heuristic Chirality Generation: Anticipations and Challenges}
\label{sec:heuristic}

As anticipated earlier fascinating physics emerge with the addition 
of a parallel magnetic field into an electric field. Whereas the electric 
field gives rise to produced pairs through the Schwinger mechanism,
the magnetic field projects the pairs' spins onto itself producing a 
net chirality. Let us begin with a cursory look at this process,
starting with the Schwinger mechanism in parallel homogeneous
fields.

Parallel homogeneous fields allow us to study a parity-violating
configuration without being encumbered by technical 
difficulties. Also, it has been reasoned such fields may give 
rise to a net chirality
through the Schwinger mechanism~\cite{Fukushima:2010vw}.
The parallel homogeneous
fields we use are in the $x^3$ direction:
\begin{equation}
   \vec{B}=B\,\hat{x}^3,\quad \vec{E}=E\,\hat{x}^3\,.
   \label{eq:homogeneous}
\end{equation}
The usage of homogeneous fields is just in that in ion-ion
collisions, related chromoelectromagnetic flux tubes are thought 
to form in the glasma~\cite{Kharzeev:2001ev,Lappi:2006fp}.

A measure of Schwinger pair production is provided
through the imaginary part of the effective action, the vacuum
non-persistence, Eq.~\eqref{eq:probability}. The 
relation describing a single particle anti-particle pair is
given by
\footnote{
Note here only the lowest order pole in 
the effective action is considered to make the  
single pair interpretation valid. Actually, the probability
of a single pair generated is given as a geometric series over all 
poles~\cite{fradkin1991quantum}, and the imaginary part of the 
effective action predicts
any number of pairs generated from the vacuum. 
See also Refs.~\cite{Cohen:2008wz,Tanji:2010eu}}
\begin{align}
   2 \text{Im} \Gamma\approx Vt\,\omega\,,
   \label{eq:nonpersistence_2}
\end{align}
where $\Gamma$ is the effective action and $V$ and $t$ are the
volume and time measures of the system. $\omega$ is the probability
that a pair is produced in a given unit space-time. For the case
of our homogeneous fields, Eq.~\eqref{eq:homogeneous},
Schwinger's formula is famously known as (see e.g., 
Ref.~\citen{Dunne:2004nc}) 
\begin{equation}
  \omega = \frac{e^2 E B}{4\pi^2}
  \coth\Bigl(\frac{B}{E}\pi\Bigr)
  \exp\Bigl(-\frac{\pi m^2}{eE}\Bigr)\,.
  \label{eq:schwinger_formula}
\end{equation}
The Landau levels are contained in the cotangent function, and
moreover in the lowest Landau level approximation (LLLA),
Schwinger's formula can be seen to resemble the non-conservation
of chirality for the chiral anomaly. This is
not coincidental we will show throughout this review. 

\begin{figure}
\centering
  \includegraphics[width=0.7\columnwidth]{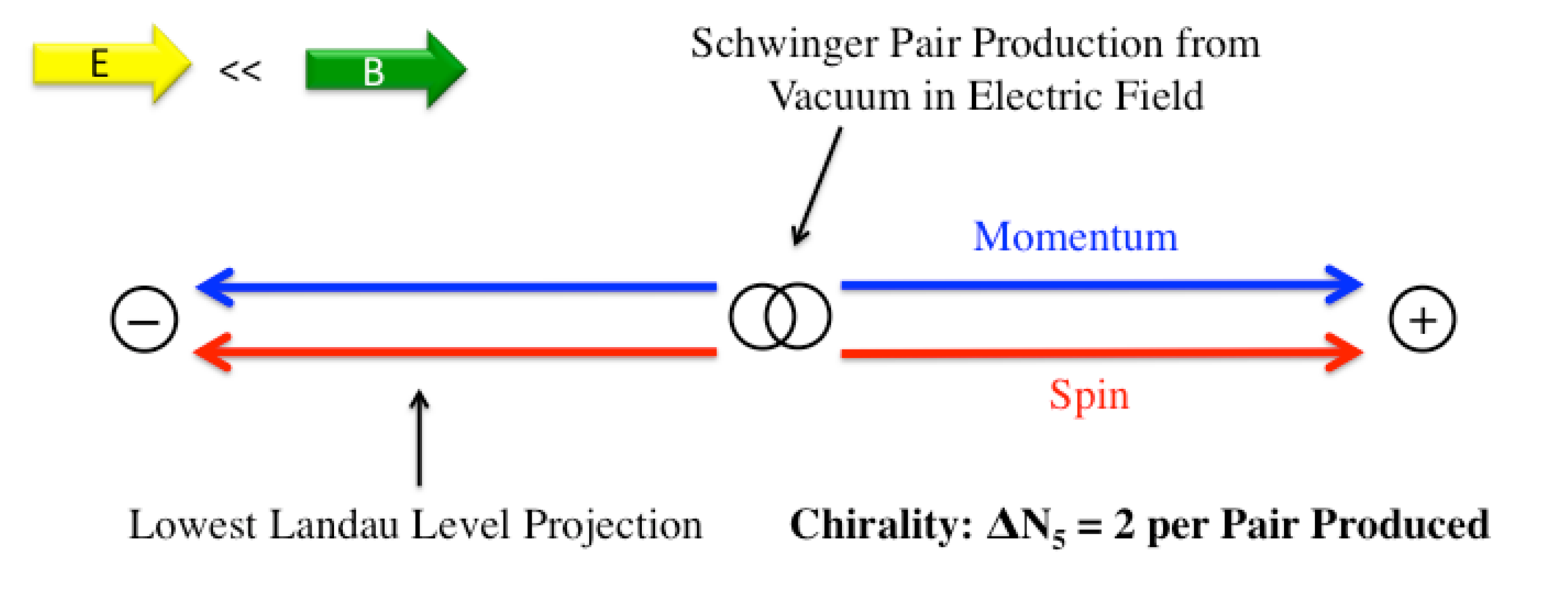}
  \caption{Cartoon of chirality production from the Schwinger mechanism.
  A LLLA is taken, and pairs of produced particles will have their spins
  aligned with the magnetic field, therefore setting up a net chirality,
  $\Delta N_5 = 2$, c.f., Fig.~\ref{fig:cme}.}
  \label{fig:chirality}
\end{figure}

The intuitive picture of chirality generation via the Schwinger
mechanism is as follows: Particle anti-particles pairs are spawned from 
the vacuum from the Schwinger mechanism. And in a strong magnetic
field parallel to the electric field, such that the LLLA may be taken,
the particles' spins will be projected to the magnetic field such that
a net chirality be generated. In Fig.~\ref{fig:chirality}, a cartoon of 
a produced pair can be seen; there a net chirality of $\Delta N_5 =2$
is produced. 
Also, one needn't assume massless fermions; with only the lowest
Landau level being occupied, an effective dimensional reduction will occur,
and chirality will be fixed. 
Let us also digress on the conventions of chirality, as is also outlined in 
Ref.~\citen{PhysRevD.78.074033}. 
Massless, or massive in a LLLA, particles with right-handed
helicity (spin and momentum are parallel) have right-handed chirality. 
Whereas anti-particles with left-handed
helicity, (spin and momentum are anti-parallel), have right-handed chirality. 
Therefore $\Delta N_5$ can be read as the total number of 
right-handed helicity particles
and anti-particles minus the total number of left-handed helicity
particles and anti-particles.

We can also benefit from a Dirac sea perspective of the chirality generation
process from pair production, as is commonly invoked to explain the
Schwinger mechanism. We did so in Fig.~\ref{fig:sea_position}, 
however, a coordinate representation was used there. Here 
we use a momentum representation. A key point in 
Fig.~\ref{fig:sea_position} is that the electric field augments the 
spectrum, enabling a traversal of the mass gap--a tunneling 
phenomenon. 
And an important aspect of the
anomaly is a QFT vacuum instability, making possible
a non-conservation of chirality~\cite{AMBJORN1983381}. 

\begin{figure}
\centering
  \includegraphics[width=0.5\columnwidth]
  {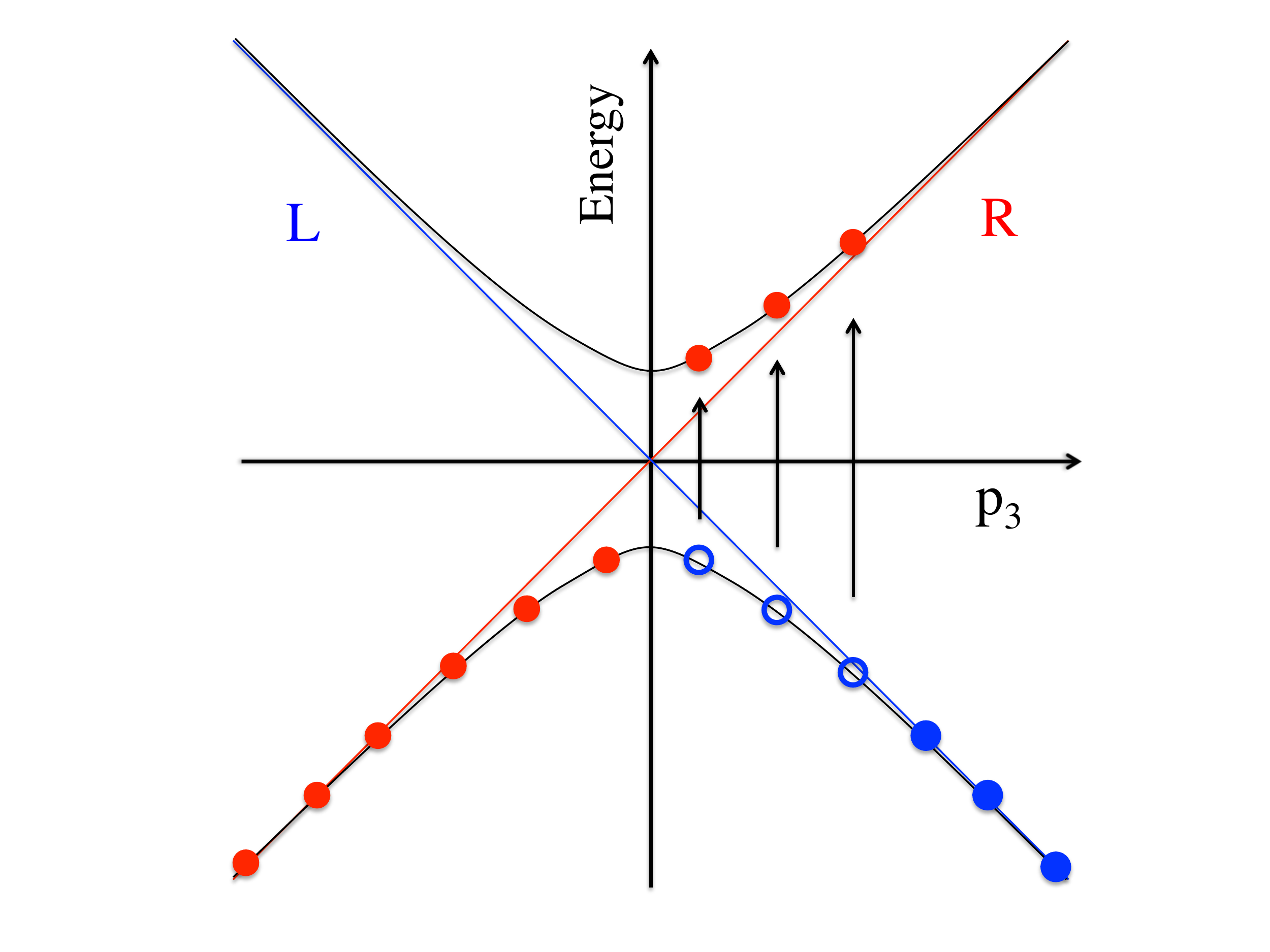}
  \caption{Dirac sea dispersion relation in momentum space. 
  Dimensional reduction for the massive fermion system is 
  made possible due to the LLLA.
  Particles tunnel from the Dirac sea leaving anti-particles in their place. Only
  right-handed particles and left-handed anti-particles are formed, setting up
  a net chirality.} 
  \label{fig:sea_momentum}
\end{figure}

The energy dispersion relation for massive fermions in parallel
fields can be seen in Fig.~\ref{fig:sea_momentum}. 
Again, a LLLA is assumed and hence we have a definite 
projection of helicity. In the electric
field a particle may tunnel from the Dirac sea, leaving an anti-particle in
its place. Then due to the strong magnetic field only particles with 
right-handed chirality and anti-particles with left-handed chirality can 
be formed; a chirality non-conservation forms as indicated by
 the axial Ward identity.
The infinite Dirac sea supplies particle non-conservation and in turn 
the anomaly through tunneling. The challenge, we will find, is in 
the determination of expectation values.
Depending on how the vacuum states are constructed, different 
physics emerges. 

Let us quantify the heuristic picture. One may expect for the 
probability density in unit time of pairs to be produced, in 
parallel fields, Eq.~\eqref{eq:schwinger_formula}, under the 
LLLA, to give rise to the following non-conservation of
chiral density:
\begin{equation}
  \omega= \frac{e^2 E B}{4\pi^2} \exp\Bigl(-\frac{\pi m^2}{e E}\Bigr) 
  \sim \frac{1}{2}\partial_0 n_5\,,
  \label{eq:schwinger}
\end{equation}
as illustrated in Ref.~\citen{Fukushima:2010vw}. Here the chiral density,
$n_5$, is the expectation value--to later be defined concretely--of the 
axial current, 
\begin{equation}
j^\mu_5 \coloneqq \bar{\psi}\gamma^\mu \gamma_5 \psi.
\end{equation}

The axial Ward identity is exact at the operator level and reads,
\begin{equation}
\partial_\mu j_5^\mu = -\frac{e^2}{16\pi^2}
\epsilon^{\mu\nu\alpha\beta}
 F_{\mu\nu}F_{\alpha\beta} + 2m\bar{\psi} i \gamma_5\psi\,.
 \label{eq:awi}
\end{equation}
Then for our field configuration, Eq.~\eqref{eq:homogeneous}, we
find the expectation value of the above becomes
\begin{equation}
  \partial_0 \langle j_5^0 \rangle = \frac{e^2 EB}{2\pi^2}
  + 2m \langle\bar{\psi}i\gamma_5 \psi\rangle\,.
  \label{eq:ward}
\end{equation}
The discrepancy stems when one performs actual calculation for the 
above. Notably, Schwinger first performed the calculation
of the pseudoscalar condensate in Ref.~\citen{Schwinger:1951nm}, 
while studying the neutral meson and proton, to find
\begin{equation}
  \bar{P}\coloneqq \langle\bar{\psi}i\gamma_5\psi\rangle
  = -\frac{e^2 EB}{4\pi^2 m}\,.
  \label{eq:pseudoscalar}
\end{equation}
What is more is that when using the above calculation in the 
axial Ward identity, Eq.~\eqref{eq:ward}, 
we find $\partial_0 \langle j_5^0 \rangle=0$! 
When compared to the heuristic picture, Eq.~\eqref{eq:schwinger},
we find an enigma:
\begin{equation}
   n_5 \neq \langle j_5^0 \rangle\,.
   \label{eq:enigma}
\end{equation}
This is valid
for any $m$, including massless fermions. It is often the case that
$m\langle\bar{\psi}i\gamma_5\psi\rangle$ is dropped for 
$m\rightarrow 0$ theories, but we can see here that the step 
is unjustified. Let us also point out, massless Abelian theories
differ from $m\rightarrow 0$ theories in that the former has
a completely shielded electric charge~\cite{Fomin:1976am}.
A resolution to the above enigma, we will demonstrate, 
can be had with an identification of vacuum states. Depending on the 
makeup of the vacuum states, expectation values
can differ markedly, both quantitatively and through 
physical interpretation.

\section{Vacuum States and Expectation Values}
\label{sec:vacuum}

\begin{figure}
\centering
  \includegraphics[width=0.5\columnwidth]{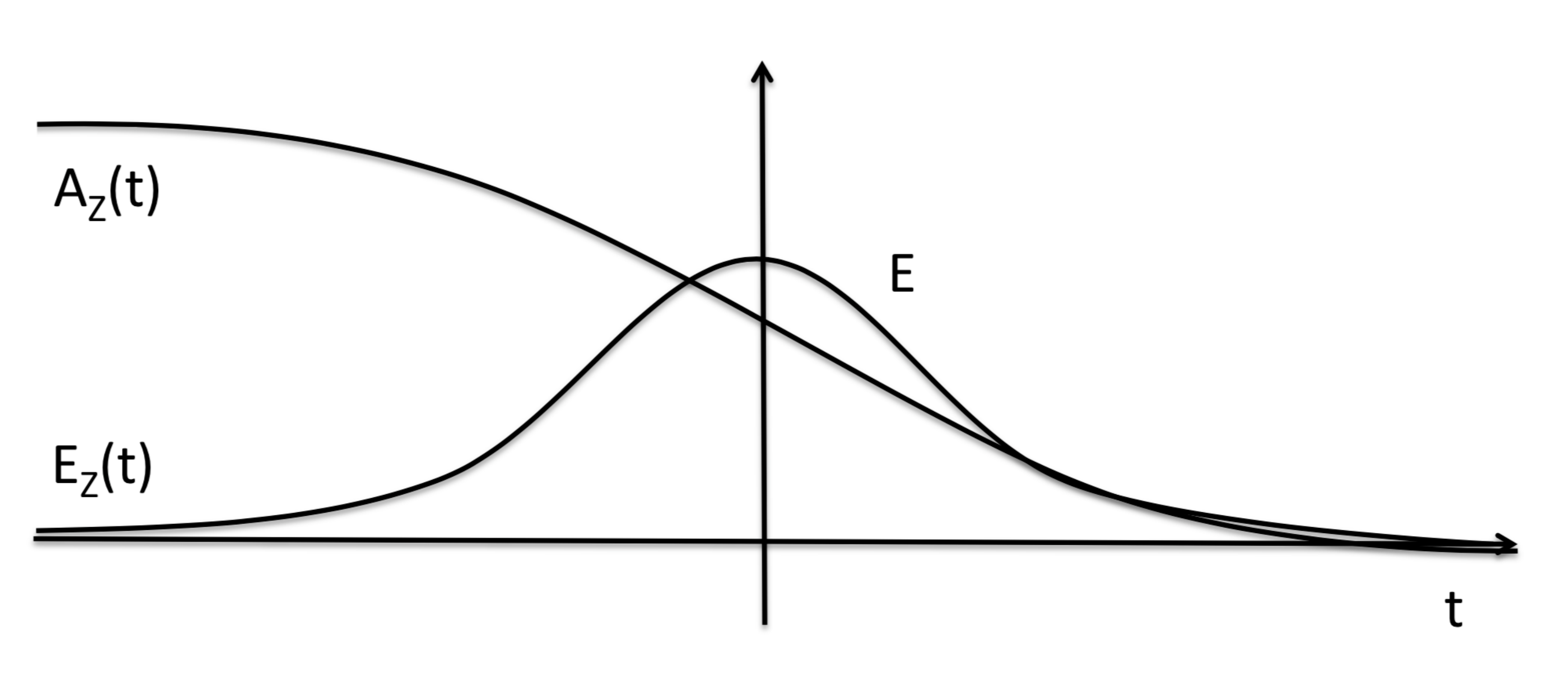}
  \caption{Sauter potential as an example of an inequivalent vacuum
   state profile. $E_z(t)=E\cosh^{-2}(t)$ and $A_z(t)=-E\tanh(t)+E$. 
   Note that while the field vanishes at $t\rightarrow \pm \infty$,
   the gauge does not.}
  \label{fig:sauter}
\end{figure}

There are implications for expectation values derived from
 fields and their gauges whose 
behavior differs at asymptotic 
times. This is even the case, for
example, for the Sauter potential~\cite{sauter}.
Note that we, 
rather, use homogeneous
fields throughout this review. Even though the field
disappears in the asymptotic limits the gauge does not;
see Fig.~\ref{fig:sauter}.
If then the background field differs at its asymptotic limits, then
the corresponding vacuum states too are affected. This 
leads to a vacuum instability, and for the case of a background
electric field, manifests itself as the Schwinger mechanism. 
The differing vacuum states are characterized as
$\instate \neq \outstate$.  Na\"ive usage of
vacuum states under a vacuum instability
 in the calculation of expectation values 
may lead to physical interpretations being marred. 
This we will show was the case
for the pseudoscalar condensate, Eq.~\eqref{eq:pseudoscalar}.
Let us emphasize, there is nothing wrong with the 
calculation leading to Eq.~\eqref{eq:pseudoscalar}. In fact,
its physical interpretation is profound, we will show. But,
how might one calculate values in accordance with our
heuristic understanding? This is accomplished by noting that
the Schwinger mechanism is an inherently out-of equilibrium
phenomenon. And as such, calculations therein can only be had with
 techniques with out-of equilibrium capacity.
The in-in formalism~\cite{fradkin1991quantum} 
provides a means. 
The formalism's usage is intuitive as well in that 
expectation values are manifestly real and coincide with
quantum mechanical expectation value definition. 

We employ
two vacuum state expectation value types throughout 
this review, both the conventional in-out and 
in-in types, which we contrast for operator, $\mathcal{O}(t)$,
as
\begin{equation}
  \langle\calO\rangle := \stateout \calO(t) \instate/c_v \,,\qquad
  \llangle\calO\rrangle := \statein \calO(t) \instate ,
  \label{eq:inout_def}
\end{equation}
where the $c_v$ is defined in Eq. (\ref{eq:cv}).
Here we emphasize again that 
the  in and out vacuum states are defined at asymptotic
times $t_{in}\rightarrow -\infty$ and 
$t_{out} \rightarrow \infty$ respectively. 

One may expand 
Dirac operators, conveniently if they wish, at the 
asymptotic times with creation and annihilation operators
acting on the respective vacuum state.
We expanded the wave function as,
\begin{align}
   \psi(x)&=\sum\limits_n a_n^{in} \phi_{+n}^{in} (x)+b_{n}^{in\,\dagger}
   \phi_{-n}^{in}(x)\nonumber\\
   &=\sum\limits_n a_n^{out} \phi_{+n}^{out}(x)
   +b_n^{out\,\dagger} \phi_{-n}^{out}(x)\,.
   \label{eq:psi_out}
\end{align} 
where $\phi_{+n}$ ($\phi_{-n}$) depicts an eigenvector of the Dirac
equation with a positive (negative) energy solution and eigenvalue,
$n$. Both 
the in and out representations are valid over all times. The creation and 
annihilation operators, for example in the in basis, act such that
\begin{equation}
   a_{n}^{in}\ket{in}= b_{n}^{in}\ket{in}
   =\bra{in}a_{n}^{in\,\dagger}=\bra{in}b_{n}^{in\,\dagger}=0\,,
   \label{eq:creation_annihilation}
\end{equation} 
with the usual anti-commutation relations applying:
$\{a_{n}^{in},a_{m}^{in\,\dagger}\}	
=\{b_{n}^{in},b_{m}^{in\,\dagger}\}=\delta_{nm}$.
One may construct a similar set for the out basis as well.

For the calculation of expectation values indicated in
Eq.~\eqref{eq:inout_def},
we introduce two useful respective 
causal propagators 
\begin{align}
  \sprop(x,y)&=i\langle T \psi(x)\bar{\psi}(y) \rangle\,,
  \label{eq:in-out} \\
  S_{in}^c(x,y)&=i\llangle T \psi(x)\bar{\psi}(y) \rrangle\,.
  \label{eq:in-in}
\end{align}
While both of the propagators satisfy a similar differential 
equation,
\begin{equation}
 -(i\slashed{D}_x -m)S^{c}_{\textrm{null},in}(x,y)=\delta(x-y),
\end{equation}
 their
boundary conditions and behavior differ.

\subsection{In-Out Propagator}
\label{sec:in-out}

Let us examine first the more conventional in-out propagator. 
As illustrated with the above arguments, expectation values
sought using the in-out propagator correspond to a matrix element
with ground states at asymptotic times, i.e.,
 $x^0\rightarrow \pm \infty$.
The meaning of such observables is fascinating in its own right and thus
we elaborate in some depth later; here in this section, however, 
we confine our attention to the derivation of the in-out propagator. 

The in-out propagator is defined from a matrix element for
asymptotic in to out states, and reads in path integral form as
\begin{equation}
\sprop(x,y)	=\int\D \bar{\psi} \D \psi \, \psi(x)\bar{\psi}(y)\,
\exp\Bigl\{i\int d^4 x'\,\bar{\psi}(i\slashed{D}-m)\psi \Bigr\}\,.
\label{eq:path_in_out}
\end{equation}
In contrast to the in-in propagator for inequivalent vacuum 
states\footnote{For equivalent vacuum states the in-out and in-in 
propagators coincide.},
the above permits a formal but simple functional representation
in proper time:
\begin{equation}
   \sprop(x,y)=\langle\text{x}| \frac{-1}{i\hat{\slashed{D}}-m}|\text{y}\rangle 
   =(i\slashed{D}_x+m)\, \langle\text{x}| \frac{1}{\hat{\slashed{D}}{}^2 
   +m^2}   |\text{y}\rangle\,.
   \label{eq:ioprop}
\end{equation}
Hats, (e.g. $\hat{\calO}$), denote operators, and are acted upon by states in 
3+1 spacetime denoted with brackets. We also stress there is a small imaginary 
piece implicit in the mass term, $m^2 \rightarrow m^2 -i\epsilon$, that is left
out for brevity. The small imaginary piece dictates the time ordering and also
guarantees convergence in the infrared limit.

The connection to Schwinger proper time~\cite{Schwinger:1951nm}
\footnote{For path integral representations of Schwinger proper time see
Refs.~\citen{Schubert:2001he,Corradini:2015tik}} is 
accomplished with a Laplace transform,
\begin{equation}
\hat{\calO}^{-1}=
i\int_0^\infty ds\,\exp(-i\hat{\calO}s)\,,
\end{equation}
with $s$ representing a proper time-like parameter. Using 
Eq.~\eqref{eq:ioprop} one can find
\begin{align}
   \sprop(x,y)&=(i\slashed{D}_x+m)\int^\infty_0 ds\, g(x,y,s)\,, 
   \label{eq:G_io} \\
   g(x,y,s)&\coloneqq i\langle\text{x}| e^{-i\hat{H}s}
   |\text{y}\rangle\,,
   \label{eq:kernel}\\
   \hat{H}&\coloneqq \hat{\slashed{D}}{}^{2}+m^2 \,,
   \label{eq:w_hamiltonian}
\end{align}
for kernel, $g$ and proper time Hamiltonian, $\hat{H}$. It is convenient
to express the kernel in its path integral form, and this is easily done by
finding the accompanying Lagrangian, $\mathcal{L}$, for the Hamiltonian,
and also through the use of the identity
\begin{equation}
   \langle x|e^{-i\hat{H}s}|y\rangle=\int_{x(0)=y}^{x(s)=x}
   \mathcal{D}x\, \mathcal{P} e^{i\int_{0}^{s}d\tau\mathcal{L}}\,.
   \label{eq:hamtopath}
\end{equation}
We can find the Lagrangian through a Legendre 
transform~\cite{Schwartz:2013pla}, where operators in 
Heisenberg notation, $\hat{\calO}(\tau)$, follow Heisenberg equations 
of motion in proper time, $\tau$, as
 \begin{equation}
\dot{\hat{\calO}}\coloneqq
\frac{d\mathcal{\hat{O}}}{d\tau}=-i[\mathcal{\hat{O}},\hat{H}]\,.
 \end{equation}
The canonical commutation relations read 
$[\hat{p}_{\mu},\hat{x}_{\nu}]=ig_{\mu\nu}$.
Then using Eq.~\eqref{eq:w_hamiltonian}, we can find the velocity as
$\dot{\hat{x}}_\mu=2(\hat{p}_{\mu}-eA_{\mu}(\hat{x}))$.
And the Lagrangian from the Legendre transformation,
$\mathcal{\hat{L}}=\hat{p}_\mu 
\frac{\partial \hat{H}}{\partial \hat{p}^\mu} -\hat{H}$, is
\begin{equation}
   \mathcal{\hat{L}}=-\frac{1}{4}\dot{\hat{x}}^{2}-eA(\hat{x})\dot{\hat{x}}
   -\frac{e}{2}F(\hat{x})\sigma-m^{2}\,,
\end{equation}
where we have used a contracted notation for the Lorentz indices, e.g. 
$F\sigma = F^{\mu\nu} \sigma_{\mu\nu}$ with 
$F^{\mu\nu}$ being the field strength tensor and 
$\sigma^{\mu\nu}:= \frac{i}{2}[\gamma^\mu,\gamma^\nu]$.
Finally the kernel in path integral form using Eq.~\eqref{eq:hamtopath} 
is identified as 
\begin{equation}
   g(x,y,s) =i\int_{x(0)=y}^{x(s)=x}\mathcal{D}x\, 
   \mathcal{P}\exp\Bigl\{i\int_0^s d\tau 
   \bigl[-\frac{1}{4}\dot{x}^2 -eA\dot{x}
   -\frac{e}{2}F\sigma-m^2\bigr] \Bigr\}\,,
   \label{eq:path_integral}
\end{equation}
where $\mathcal{P}$ indicates time-ordering for proper time $\tau$.
This is the kernel of the worldline path integral
~\cite{Feynman:1950ir,Feynman:1951gn}, and while the above 
form is valid for any background QED field, 
at this point let us restrict our attention to the case of parallel 
electric and magnetic fields, Eq.~\eqref{eq:homogeneous}. 
The in-out propagator in homogeneous parallel fields is a 
well-known expression~\cite{Schwinger:1951nm,Schwartz:2013pla}, 
however steps worked through 
in its derivation will aid in later discussions. 

For homogeneous fields we can  factorize the kernel into 
both a spin factor $\Phi(s)$
and boson path integral $b(s)$ such that they are connected
through proper time as in
\begin{equation}
g(s) = b(s)\Phi(s)\exp(-im^2s),
\end{equation}
with
\begin{align}
   b(x,y,s) &\coloneqq\int_{x(0)=y}^{x(s)=x}\mathcal{D}x\,
    \exp\Bigl\{ i\int_0^s d\tau 
   \bigl[-\frac{1}{4}\dot{x}^2 -eA\dot{x}\bigr] \Bigr\}\,,
   \label{eq:bpath}\\
\Phi(s) &\coloneqq \mathcal{P} \exp\Bigl\{
 -i \int^s_0 d\tau  \frac{e}{2} F\sigma
  \Bigr\}\,.
  \label{eq:spin_factor}
\end{align}
We first address the spin factor. For our choice of fields in the $x^3$
direction and with the use of Weyl gamma matrices, 
Eq.~\eqref{eq:weyl_gamma}, the spin factor
takes a diagonal form. The path ordering is negated simplifying matters.
Here we represent the spin factor using gamma matrices as
\begin{equation}
   \Phi(s)=[\cos(eBs)+i\sin(eBs)\sigma^{12}]
   \times[\cosh(sEs)+\sinh(eEs)\gamma_{5}\sigma^{12}]\,,
   \label{eq:spin_factor2}
\end{equation}
with $\sigma^{12}=\diag[1,-1,1,-1]$. Then all that is needed to solve
the in-out propagator is to determine the boson path integral.

In homogeneous fields the boson path integral, Eq.~\eqref{eq:bpath}, 
has an exact solution in steepest descents owing to the 
quadratic form of coordinates in the action.
We use the Fock-Schwinger gauge,
\begin{equation}
A_\mu(x)=-\frac{1}{2}F_{\mu\nu}x^\nu.
\end{equation}
We can evaluate the path integral though steepest descents; we 
expand $x$ about the classical path, $x^{cl}$, such that 
$x_\mu(\tau)=x_\mu^{cl}+\eta_\mu(\tau)$.
for small fluctuations, $\eta$. The fluctuations disappear at
the endpoints, $\eta(0)=\eta(s)=0$. Then for the worldline
action, 
$S_b=\int_{0}^{s}d\tau[-\frac{1}{4}\dot{{x}}^2-eA\dot{{x}}]$,
one can find Eq.~\eqref{eq:bpath} becomes
\begin{align}
  b&(x,y,s)=e^{iS_b(x^{cl})}\mathcal{F}\, ,\\
  \mathcal{F}&\coloneqq \int \mathcal{D}\eta \,
  \exp\Bigl\{i\int_0^sd\tau\bigl[-\frac{1}{4}\dot{{\eta}}^{2}
  +\frac{1}{2}\eta eF\dot{\eta}\bigr]\Bigr\}\,.
  \label{eq:fluc_pre}
\end{align}

The classical equation of motion for the boson worldline action is
simply the Lorentz force equation 
\begin{equation}
  \ddot{x}^{cl\, \mu}(\tau)=2eF_{\hspace{0.3em} \nu}^{\mu}
  \dot{{x}}^{cl\, \nu}(\tau) \,,
  \label{eq:lorentz} 
\end{equation}
with solution,
$\dot{x}^{cl\, \mu}(\tau)=
 [e^{2eF\tau}]_{\hspace{0.3em} \nu}^{\mu}
  \dot{{x}}^{cl \, \nu}(0)$.
Then for the displacement
\begin{equation}
z\coloneqq x-y\,,
\end{equation}
and taking note of the boundary conditions for the boson
path integral one can find that
$\int_0^s d\tau\dot{x}^{cl\, \mu} (\tau)=z^\mu$.
Also it can be found that
$(e^{2Fs}-1)_{\hspace{0.3em} \lambda}^{\mu} \dot{x}^{cl\, \lambda}
(0)=2F_{\hspace{0.3em} \lambda}^{\mu}z^{\lambda}$. Last, 
using the above relationships one can find the for the classical
worldline action 
\begin{align}
  \varphi &\coloneqq S_b(x^{cl}) 
  =\frac{1}{2} x \,eF \,y
  -\frac{1}{4}z \,\coth(eFs)\,eF \,z\\
  &=\frac{1}{2} x \,eF \,y +\frac{1}{4}\bigl[ (z_3^2-z_0^2)eE\coth(eEs) 
   +(z_1^2+z_2^2)eB\cot(eBs)\bigr]\,.
   \label{eq:class_action}
\end{align}
Let us point out that all the gauge dependence resides in the $x\,eF\,y$ term, 
and also that after application of the covariant derivative acting on the 
kernel, the classical worldline action vanishes as $x\rightarrow y$.

One may calculate the fluctuation prefactor, Eq.~\eqref{eq:fluc_pre},
by expanding about Fourier modes, i.e.:
\begin{equation}
   \eta_{\mu}(\tau)=a_{\mu_{0}}+\sum\limits _{n=1}^{\infty}\Bigl[
   a_{\mu \,n}\cos\Bigl(\frac{2\pi n\tau}{s}\Bigr)
   +b_{\mu\,  n}\sin\Bigl(\frac{2\pi n\tau}{s}\Bigr)\Bigr]\,.
\end{equation}
After some steps, and equipped with the free field solution, 
\begin{equation}
\int \mathcal{D} \eta \,e^{ i\int_0^sd\tau
[-\frac{1}{4}\dot{\eta}^2] }=-i / (4\pi s)^2,
\end{equation}
it can be 
found the fluctuation prefactor becomes
\begin{equation}
   \calF=-i\frac{e^2EB}{(4\pi)^2}\sin^{-1}(eBs)\sinh^{-1}(sEs)\,.
\end{equation}

Finally, we may gather all the terms to find the kernel,
Eq.~\eqref{eq:kernel}, as
\begin{equation}
   g(x,y,s)=\frac{e^2EB}{(4\pi)^{2}}
   \frac{\exp[-im^2s+i\varphi(x,y,s)]}{\sin(eBs)\sinh(eEs)}\Phi(s)\,,
   \label{eq:ker_final}
\end{equation}
with spin factor given in Eq.~\eqref{eq:spin_factor2}. Using
proper time methods we have at our disposal a wealth of physics
described in a compact expression. 
All the physics of the Schwinger mechanism is contained in
the kernel. Let us illustrate that by making the connection to
Schwinger's formula, Eq.~\eqref{eq:schwinger_formula}. The 
effective action may be expressed as
\begin{align}
\Gamma[A]&=-i\Tr\ln(i\slashed{D}-m)\\
&=\frac{1}{2} \tr \int d^4 x\int^\infty_0 \frac{ds}{s}\,g(x,x,s)\,.
\end{align}
Then evaluating for the imaginary part, taking only the contribution
of the lowest order pole at $s=-\pi / eE$, and using 
Eq.~\eqref{eq:nonpersistence_2},
one can find Schwinger's formula, Eq.~\eqref{eq:schwinger_formula}.
Also, all the spin structure of the system is contained in the 
spin factor. And last, all the Landau levels are kept in the 
$\cot (eBs)$ functions. The kernel and the in-out propagator
are exact to one loop. For extensions to QCD see 
Ref.~\cite{Hattori:2020guh}. The in-in propagator may be cast in a 
similar form, in fact with just a modification to the proper time
integral in $s$.

\subsection{In-In Propagator}
\label{sec:in-in}

The proper time representation of the in-out propagator is compact
and it would be advantageous to do the same for the in-in propagator.
This has fortunately been accomplished by Fradkin et. al. in 
Ref.~\citen{fradkin1991quantum}, and they find the only modification
of which in comparison to the in-out case is an augmentation of
the proper time integral. 

\begin{figure}
\centering{
  \includegraphics[width=0.55\columnwidth]{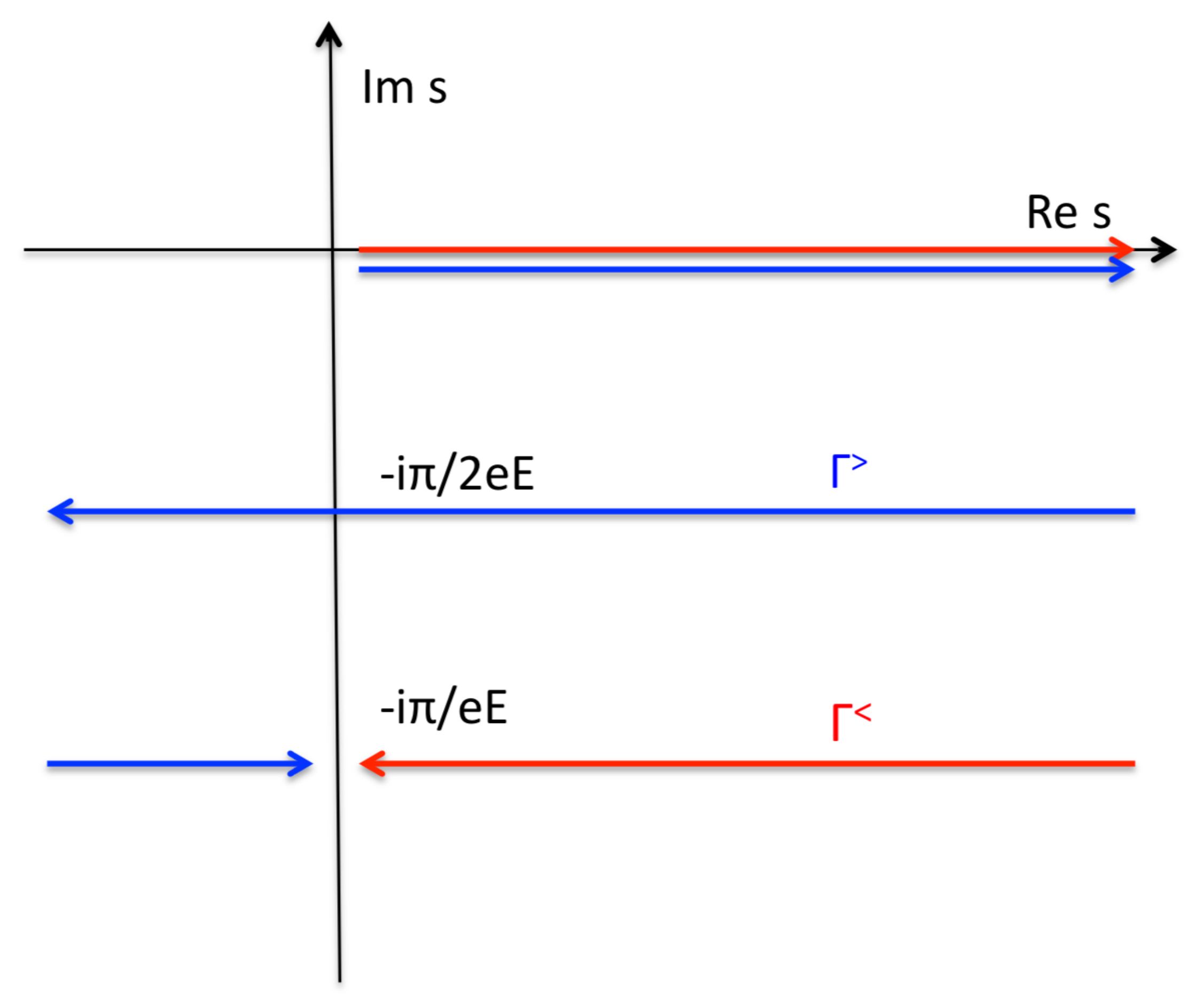}}
  \caption{In-in propagator contours in proper time, $s$. 
  $\Gamma^>$ and $\Gamma^<$ are valid for positive 
  and negative $z_3$ respectively. Contours near the real axis 
  lie slightly below it.}
  \label{fig:contours}
\end{figure}

There it is found the in-in propagator can be
cast into a Schwinger proper time representation, where the 
in-out contour has been subsumed into the overall contour, as
\begin{align}
   \Sprop(x,y)&=(i\slashed{D}_{x}+m) \int_{in}ds\, g(x,y,s) 
   \label{eq:S_ii} \\
   \int_{in}ds&\coloneqq\Bigl[\theta(z_{3})\int_{\Gamma^{>}}ds
   +\theta(-z_{3})\int_{\Gamma^{<}}ds\Bigr]\,.
   \label{eq:ii_prop_final}
\end{align}
The contours are given in Fig.~\ref{fig:contours}. 
We have 
acquired Heaviside theta function arguments in the proper time
integral and kernel. The arguments are important in that they 
give rise to the real-time dependence in our out-of equilibrium
formulation. We further illustrate their
nature with concrete examples in the coming sections. 

Recently, we have found the deep connection between 
the in-in propagator in Eq. (\ref{eq:ii_prop_final})
and famous Schwinger-Keldysh (closed-time-path) 
formalism~\cite{Schwinger:1960qe,Baym:1961zz}.  
We can also derive the  Eq. (\ref{eq:ii_prop_final})
through the analyzing the Bogoliubov coefficients of in and out 
states~\citen{fradkin1991quantum}.  
We will present these results somewhere else.
We have introduced both the in-out and in-in formalisms,
and now let us examine their differences and characteristics.

\subsection{In-Out and In-In Expectation Values}
\label{sec:expectation}

Let us consider a Wick rotation of the Lagrangian, e.g., as provided
in Eq.~\eqref{eq:cv}, such that $x^0 = -i x^4$. Naturally, this
describes a Euclidean QFT at zero temperature, with action
$\int^\beta_0 dx^4\int d^3x[\bar{\psi}(i\slashed{D}-m)\psi]$,
and hence in equilibrium. And thus in-out observables should correspond to
a Euclidean equilibrium picture. Furthermore, owing to the periodicity
of Euclidean time, the Euclidean partition function and hence its
vacuum states too ought to follow periodicity, i.e.,
$\bra{x^4_{in} = 0} = \bra{x^4_{out} = \beta}$. In spite of the
above straightforward arguments, an interpretation of Euclidean 
equilibrium for in-out observables has subtleties. 

A merit of the in-in or Schwinger-Keldysh formalism is the 
guarantee of real observables for single bilinear fermion fields provided
through the Hermiticity of their construction. This, however, is not
the case for in-out observables; there one can find
$\langle \bar{\psi} \mathcal{O} \psi \rangle 
\neq \langle \bar{\psi} \mathcal{O} \psi \rangle^*$, even 
for certain Hermitian $\calO$. 
And we will show with a concrete example 
imaginary pieces can reside in in-out observables. This problem
is present in cosmological applications as well, where an in-out 
construction may give way to a complex metric, making physical 
interpretation challenging~\cite{PhysRevD.33.444}. 
The source of the problem stems from a Wick rotation under
an electric field. Strictly speaking, a Euclidean QFT is defined 
under $U_A(1)$ with all fields real--and hence complex in Minkowski
space. In our case, we began with all real fields in Minkowski 
space then after a Wick rotation one would find complex fields in
Euclidean space; thus enlarging the gauge group. Indeed, an
imaginary electric field in Euclidean space is generally utilized in the
study of Schwinger pair production~\cite{Dunne:2005sx,Affleck:1981bma}. 
And more generally the sign problem, as is readily the case in
a Euclidean metric, is a necessary ingredient for Schwinger pair
production to occur. Despite the above reasoning, for the most physically 
relevant observables, outlined below, no imaginary piece is found and
therefore the interpretation of Euclidean equilibrium holds. And for any
case, that the in-out formalism not predict any produced pairs in the out
state always holds. Having both the in-out and in-in propagator at 
our disposal, let us proceed with the evaluation of chiral related 
expectation values.

\section{Axial Ward Identity}
\label{sec:awi}

Having determined the importance of vacuum states in the determination
of expectation values, and their (out-of) equilibrium nature, let us
proceed with concrete calculations. We begin with the 
enigma, Eq.~\eqref{eq:enigma}, or rather the axial Ward identity,
Eq.~\eqref{eq:ward}. We will illustrate how the controversy is
solved using Schwinger proper time methods both in and out-of
equilibrium as discussed in previous sections. The determination
of the axial Ward identity in homogeneous fields is entirely
dictated by the pseudoscalar condensate term. This is because 
$\langle \epsilon^{\mu\nu\alpha\beta}F_{\mu\nu}F_{\alpha\beta} \rangle
=\llangle \epsilon^{\mu\nu\alpha\beta}F_{\mu\nu}F_{\alpha\beta} \rrangle
=\epsilon^{\mu\nu\alpha\beta}F_{\mu\nu}F_{\alpha\beta}$. 

We first address the equilibrium or in-out pseudoscalar condensate.
The result was written above without proof in 
Eq.~\eqref{eq:pseudoscalar}. Here,
let us examine the quantity in the context of Schwinger proper time.
And, indeed Schwinger was the first to examine the pseudoscalar
condensate through such means~\cite{Schwinger:1951nm}.
Using the in-out propagator, Eq.~\eqref{eq:ioprop}, we have the 
compact expression for the pseudoscalar condensate in 
parallel homogeneous fields, Eq.~\eqref{eq:homogeneous},
\begin{equation}
  \bar{P} \coloneqq \langle\bar{\psi}i\gamma_5\psi\rangle
  = -\lim_{y\to x}\tr[\gamma_5 \sprop(x,y)]\,.
  \label{eq:io_pseudoscalar_setup}
\end{equation}
Evaluation of the above can be readily done. Let us begin
by noting that the portion of the covariant derivative, $\slashed{D}$
acting on the kernel, $g$, in $\sprop$ vanish owing to the
fact that an odd number of gamma matrices vanish under a trace. 
In fact, we will find more generally that this term vanishes for 
point split expectation values due to translational symmetry,
i.e., $g(x,y)=g(x-y)$. For the remaining mass dependent term in the 
pseudoscalar condensate we also find a cancellation from the
term of the spin factor, after taking the Dirac trace, and the
boson path integral fluctuation term. The remaining form reads
\begin{align}
  \bar{P} = -\lim_{y\to x}4i\frac{m e^2 EB}{(4\pi)^2} 
  \int_0^\infty ds\, e^{-im^2 s+i\varphi(x,y,s)}
    = -\frac{e^2 EB}{4\pi^2m}\,.
    \label{eq:pseudo_eq}
\end{align}
Even despite the essential singularity in $\varphi$, after taking the
$x\rightarrow y$ limit all terms with $z$ dependence vanish; we 
will show this step shortly. Using the above and Eq.~\eqref{eq:ward},
one can find that the axial Ward identity
\begin{equation}
   \partial_0 \bar{n}_5 \coloneqq \partial_0 
   \langle\bar{\psi}\gamma^0\gamma_5\psi\rangle=0\,,
   \label{eq:aa_eq}
\end{equation} 
predicts a conservation of chirality for any mass. It is quite 
astonishing that this should be the case. We will find that the enigma
and the above relationship are resolved using an in-in, or
out-of equilibrium, formalism. And thus, the chiral anomaly persists
as expected; see Eq.~\eqref{eq:schwinger}. However, 
we find here in Euclidean equilibrium 
no such non-conservation, suggesting the anomaly only exist
out-of equilibrium. One may anticipate such a scenario in the context
of a condensed matter system for the CME, a close relative of the 
anomaly. There the disappearance of the CME in equilibrium, but 
its reemergence out-of equilibrium is well-known~\cite{Yamamoto:2015fxa}. 
And the same phenomenon is echoed here for the anomaly. 
One can see why Eq.~\eqref{eq:aa_eq} should hold for the massless
case: Topological properties are independent of a $\theta$ term and hence
a nonzero topological charge or net chirality would not be expected.

Eq.~\eqref{eq:aa_eq} is valid for any mass and thus the 
pseudoscalar term should always be kept, even for small masses
in QED and QCD. However, one may discover the effects of a
mass on the axial Ward identity through the use of 
nonequilibrium techniques. Also, in doing so, we can resolve the enigma
and show the dependence of the chiral anomaly on the 
Schwinger mechanism. To reiterate, in-out, or Euclidean equilibrium,
expectation values predict no pairs of particles in the out state
generated via the Schwinger mechanism, whereas in-in, or 
out-of equilibrium, expectation values predict any number of pairs.

In analogy to the in-out case, Eq.~\eqref{eq:io_pseudoscalar_setup}, 
let us directly 
calculate the out-of equilibrium pseudoscalar condensate using 
the in-in propagator, Eq.~\eqref{eq:ii_prop_final},
\begin{align}
  P &\coloneqq \llangle\bar{\psi}i\gamma_5\psi\rrangle
  = -\lim_{y\to x}\tr[\gamma_5 \Sprop(x,y)] \nonumber\\
  &= -\lim_{y\to x}i\frac{m e^2 EB}{4\pi^2} 
  \Bigl[\theta(z_3)\int_{\Gamma^>}+\theta(-z_3)\int_{\Gamma^<}\Bigr]
   ds\, e^{-im^2 s+i\varphi(x,y,s)}\,, \label{pseudo_scalar_01}
\end{align}
where we have repeated similar steps as were taken in the Euclidean
equilibrium case above. One may actually evaluate the above for either 
$z_3 \rightarrow \pm 0$ and hence either $\Gamma^>$ 
or $\Gamma^<$, and this is due to the fact that the pseudoscalar
condensate is unaffected by a point-splitting scheme. We elect to 
use conventions
as written in Sec.~\ref{sec:introduction}. Then one may deform
the contours to obtain for the in-in pseudoscalar 
condensate~\cite{Copinger:2018ftr}, 
 \begin{align}
 P &= - 4i\frac{m e^2 EB}{(4\pi)^2}
 \Bigl[ \int^\infty_0 ds-\int^{\infty-i\frac{\pi}{eE}}_{-i\frac{\pi}{eE}}ds
 \Bigr]\, e^{-im^2 s} \nonumber\\
 &= -\frac{e^2 EB}{4\pi^2 m} 
 \Bigl[1 - \exp\bigl(-\frac{\pi m^2}{eE} \bigr)\Bigr]\,,
  \label{eq:Pin}
\end{align}
in agreement with the heuristic expression, Eq.~\eqref{eq:schwinger}.
Note that the rigorous way to evaluate Eq. \eqref{pseudo_scalar_01}
is to integrate
over $\Gamma^>$ and $\Gamma^<$ first and then take the 
$z_3 \rightarrow 0$ limit as we will show in a later calculation of 
$j^\mu_5$.

Chirality has been generated through the Schwinger mechanism.
Likewise, using Eq.~\eqref{eq:ward}, one can find for the 
in-in out-of equilibrium axial Ward identity,
\begin{equation}
   \partial_0 n_5\coloneqq \partial_0 
   \llangle\bar{\psi}\gamma^0\gamma_5\psi\rrangle=
   \frac{e^2 EB}{2\pi^2}  
   \exp\bigl(-\frac{\pi m^2}{eE}\bigr)\,.
   \label{eq:awi_ii}
\end{equation}
We have recovered the chiral anomaly, and also shown its dependence
on mass. We also find that only the Schwinger mechanism has contributed
to the non-conservation of chirality. We also point out that
Eq.~\eqref{eq:Pin} was also inferred from the axial Ward identity in 
Ref.~\citen{Warringa:2012bq}.

It is also interesting to draw the connection between the proper time formalism
and the Fujikawa~\cite{Fujikawa:1979ay} method, so we digress here.
The Fujikawa method entails that the anomaly arises from the QFT path
integral measure after performing a chiral rotation. The method predicts
the anomaly despite usage of massless fermions. 
Important in the Fujikawa method 
is the careful regularization of the functional trace of $\gamma_5$. And,
in fact, this same heat-kernel regularization process is present in the Schwinger
proper time construction: It is the ultraviolet, or small $s$, limit. 

\begin{figure}
\centering{
  \includegraphics[width=0.5\columnwidth]{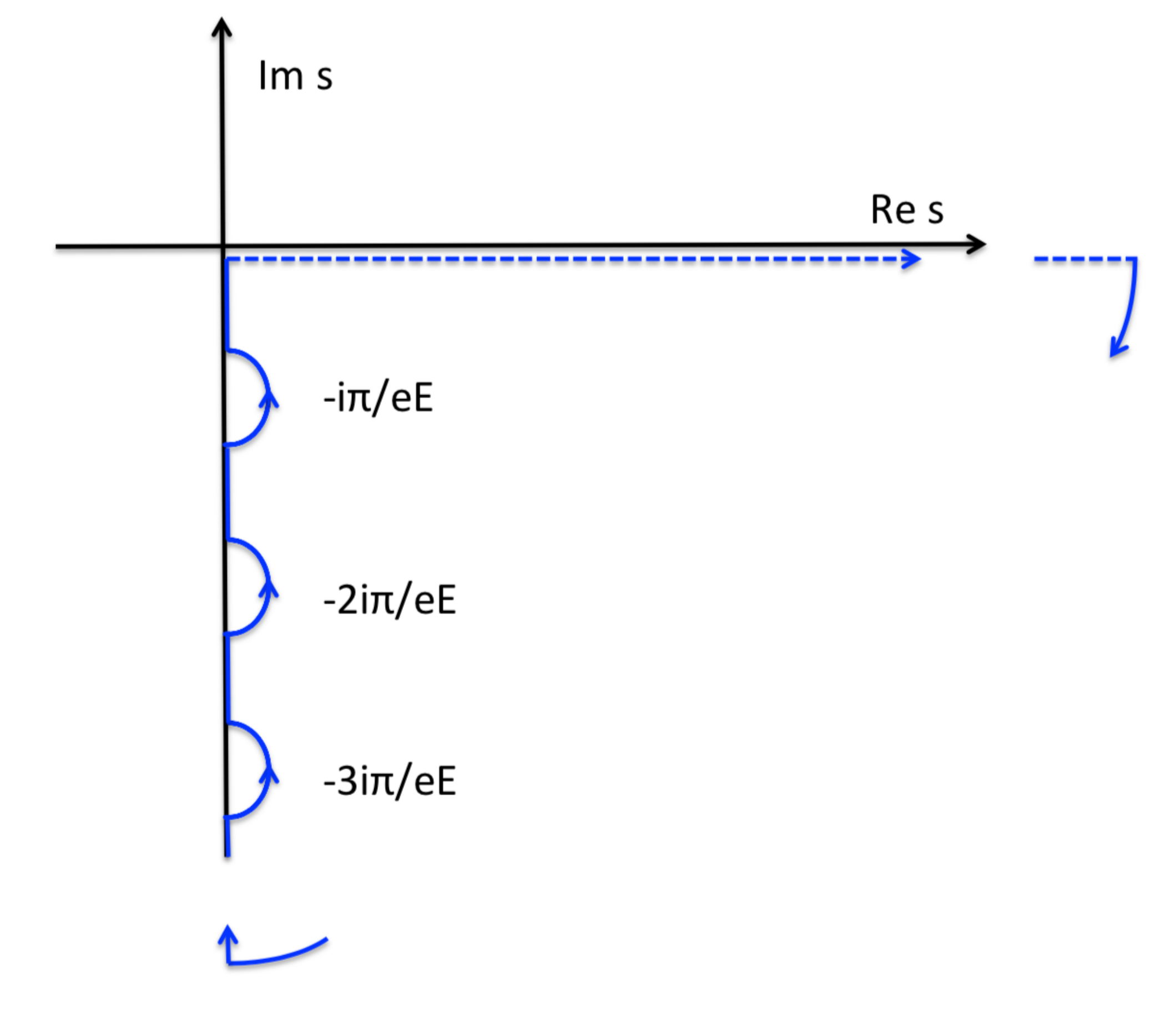}}
  \caption{Proper time contour rearrangement using Cauchy's integral 
  theorem to imaginary values. Convergence is provided by the 
  $m^2 -i\epsilon$ term.
  Singularities can be found at $s=-i\frac{n \pi}{eE}$ for $n=1,2,...$.}
  \label{fig:imag_contour}
\end{figure}

It is instructive to confirm our previous results on the axial Ward identity
by directly calculating the chiral density. Doing so, we will find, 
provides insight into the real-time nature of our out-of equilibrium
observables. As before, however, let us address the Euclidean equilibrium
case first; the chiral current, with $\bar{n}_5$ being the density, is
in proper time notation
\begin{equation}
  \bar{j}_5^\mu \coloneqq \langle\bar{\psi}
  \gamma^\mu \gamma_5\psi\rangle
  = i \lim_{y\to x}\tr[\gamma^\mu \gamma_5 \sprop(x,y)]\,.
\end{equation}
Such a term we can show for our fields, Eq.~\eqref{eq:homogeneous},
vanishes, and is therefore in agreement with Eq.~\eqref{eq:aa_eq}.
Noting again that the trace of an odd number of Dirac matrices
vanishes we can see that only the covariant derivative piece remains,
\begin{equation}
   \bar{j}_5^\mu = -i\lim_{y\to x}\tr[\gamma^\mu 
   \gamma_5\,\slashed{D}_x \int_0^\infty ds\, g(x,y,s)]\,.
   \label{eq:j5_trace}
\end{equation}
It is known that in homogeneous fields such a term should be zero;
see e.g. Refs.~\citen{fradkin1991quantum,PhysRevD.78.045017},
and this is because of translational symmetry. However, let us show
why it should be the case. Allowing the covariant derivative act
on the kernel one can find
\begin{align}
   \slashed{D}_x g(x,y,s)&=(\partial_\mu -\frac{i}{2}eF_{\mu\nu}x^\nu)
   \gamma^\mu g(x,y,s) \nonumber\\
   &=-\frac{1}{2}\Bigl[i eF_{\mu\nu} +(\coth(eFs)eF)_{\mu\nu}\Bigr]
   z^\nu \gamma^\mu g(z,s)\,,
   \label{eq:cov_derivative}
\end{align}
and hence a factor $z$ is present. Then so long as the kernel be analytic
as $x\rightarrow y$, Eq.~\eqref{eq:cov_derivative} should go to zero 
as $x\rightarrow y$. Outside of the singularities this will clearly be the 
case, however near the singularities we expand about the poles and then
take the limit. Note, we have defined the Schwinger proper time contour
so that it lies slightly below the real axis. 
The singularities in the kernel, $g$, can be seen in
Fig.~\ref{fig:imag_contour}; where we have rotated the contour
to imaginary $s$.
There are essential singularities at $-i\frac{n \pi}{eE}$ for $n=1,2,...$
in $\varphi$ encased in semicircle contours. We expand about the poles
and apply the following residue formula for pole $n$ for their treatment:
\begin{equation}
   -i\pi \text{Res}\Bigl(g,-i\frac{n \pi}{eE}\Bigr)=\frac{-i\pi}{(n-1)!}
   \lim\limits_{s\rightarrow 0}
   \frac{d^{n-1}}{ds^{n-1}}\Bigl[\Bigl(s+\frac{in\pi}{eE}\Bigr)g(s)\Bigr]\,.
   \label{eq:residue}
\end{equation}
However, with the application of the covariant derivative and upon taking
the $x\rightarrow y$ limit one can find 
\begin{equation}
   \lim\limits_{y\rightarrow x} \slashed{D}_x \, 
   \text{Res}\Bigl(g,-i\frac{n \pi}{eE}\Bigr) =0\,.
   \label{eq:res_D}
\end{equation}
Therefore, we find that the Euclidean equilibrium chiral
density, $\bar{j}_5^\mu$, vanishes and thus is
in agreement with Eq.~\eqref{eq:aa_eq} and the vanishing of the anomaly
in equilibrium. However, as anticipated earlier the out-of equilibrium
chiral density does not vanish.

\begin{figure}
\centering{
  \includegraphics[width=0.3\columnwidth]{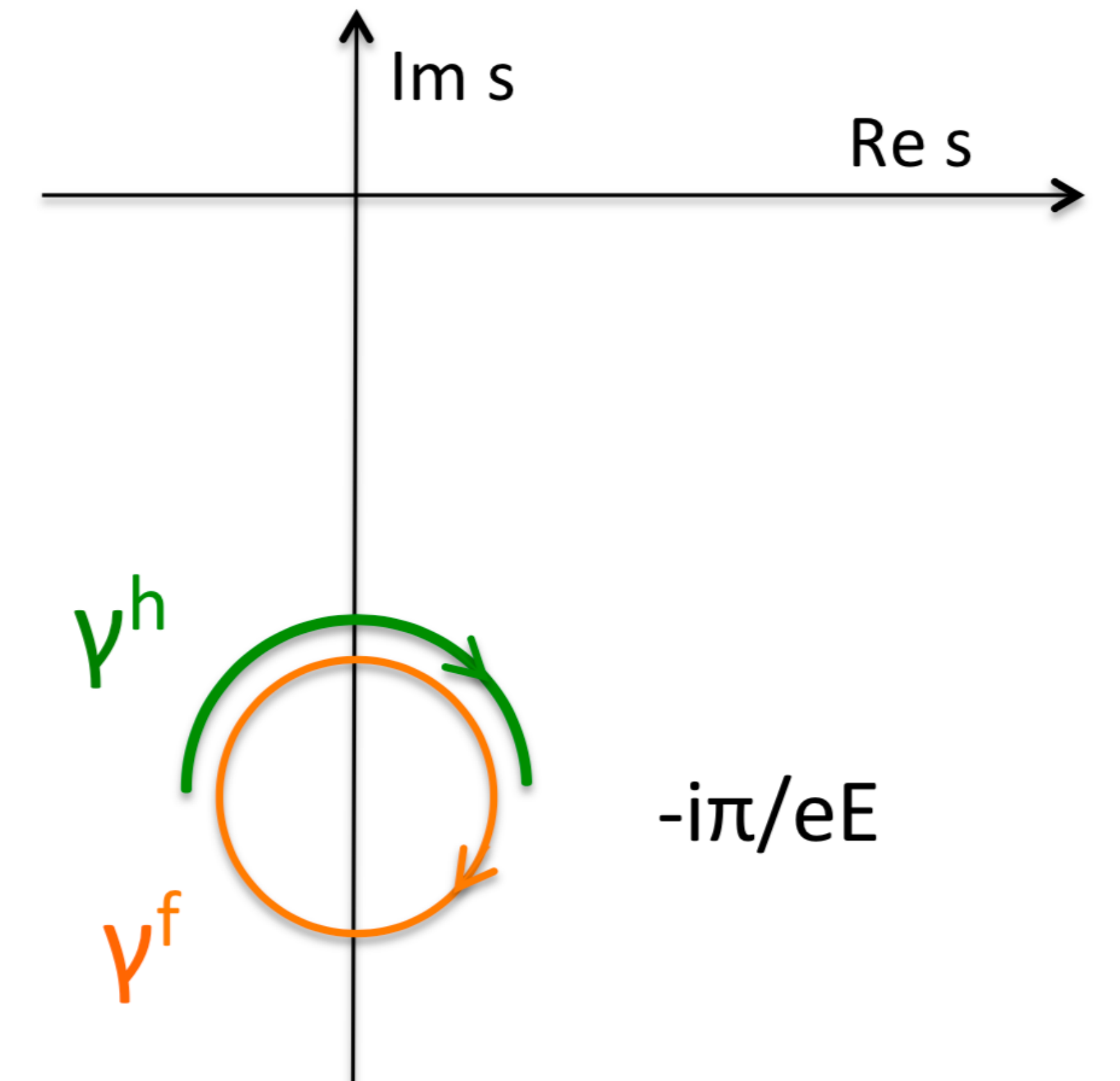}}
  \caption{Various contours in proper time, $s$, used throughout.}
  \label{fig:fluc_contours}
\end{figure}

The in-in, or out-of equilibrium, chiral density is 
\begin{equation}
  j_5^\mu \coloneqq \llangle\bar{\psi}
  \gamma^\mu \gamma^5\psi\rrangle
  = i \lim_{y\to x}\tr[\gamma^\mu \gamma_5 S^c_{in}(x,y)]\,.
\end{equation}
Observing that the trace of an odd number of gamma matrices vanishes,
\begin{equation}
   j_5^\mu = -i\lim_{y\to x}\tr[\gamma^\mu 
   \gamma_5\,\slashed{D}_x \int_{in} ds\, g(x,y,s)]\,.
   \label{eq:j5_in_trace}
\end{equation}
We find the key difference with the in-in formalism and 
Eq.~\eqref{eq:j5_trace} is an augmentation of the proper time
contour to include $\theta(\pm z_3)$. Whereas before,
Eq.~\eqref{eq:res_D}, we found that the kernel, after being acted on by 
a covariant derivative and $x\rightarrow y$ limit, vanished, here
we find real-time dependence arises from a phase space factor.
Let us rearrange the proper time contours given above for both 
$\Gamma^<$ and $\Gamma^>$ as done in Fig.~\ref{fig:imag_contour};
what remains is a semicircle contour about $s=-i\pi/eE$. We denote
the semicircle contour as $\int_{\gamma^h}$; see 
Fig.~\ref{fig:fluc_contours} for the semicircle,
as well as other contours used later.
In contrast to the Euclidean equilibrium case, we find certain residues do not 
disappear after taking the $x\rightarrow y$ limit. Let us
illustrate that fact with a sample encountered integral:
\begin{equation}
   I_h =  \int_{\gamma^h} ds\,
   e^{-im^2s +i \varphi(x,y,s)}\coth(eEs)\,.
   \label{eq:I_h}
\end{equation}
We first shift the proper time argument such that 
$s\rightarrow s'+\frac{i\pi}{eE}$ and keep only leading terms in $z$ for 
small $z$ in the integrand, to find
\begin{align}
   I_h&\approx e^{-\frac{m^2\pi}{eE} }
   \int_{\gamma^h+i\frac{\pi}{eE}} 
   \frac{ds}{eEs} e^{-\frac{i}{4s}(z_0^2-z_3^2)}\\
   &=e^{-\frac{m^2\pi}{eE} }
   \int^\infty_{-\infty}\frac{d\eta}{eE\eta} 
   e^{-\frac{i}{4}(z_0^2-z_3^2)\eta}
   =-\frac{2\pi i }{eE} \theta(z_3^2-z_0^2)
   e^{-\frac{m^2\pi}{eE} }\,
   \label{eq:I_h}
\end{align}
where in the second step we have made the change of variables,
$\eta=1/ s$, leading to a Heaviside function. 
One may perform a similar set of computations to find 
integrals without the $\coth (eEs)$ factor in the integrand,
from the spin factor, Eq.~\eqref{eq:spin_factor2}, vanish in the
$x\rightarrow y $ limit. Let us also mention in passing that one could 
also use the relation for the singularity, 
$\int_{\gamma_h}ds\,g
=\theta(z_3^2-z_0^2)\int_{\gamma_f}ds\,g$, 
and the residue formula, Eq.~\eqref{eq:residue} to find 
Eq.~\eqref{eq:I_h}; see Fig.~\ref{fig:fluc_contours}. Returning
to $j^\mu_5$, let us take the Dirac trace and keep terms with 
a $\cosh(eEs)$ in the spin factor, 
(and hence $\coth(eEs)$ factor when combined with 
the boson path integral factor), as dictated above. We find for
the Dirac trace
\begin{align}
   &\tr\gamma_0\gamma_5\slashed{D}\exp(-\frac{i}{2}eF\sigma s)
   =D_\nu \Bigl[ i\sin(eBs)\cosh(eEs)
   \tr(\gamma_0\gamma_5 \gamma^\nu \sigma^{12})\nonumber \\
   &-\frac{1}{2}\sin(eBs)\sinh(sEs)\tr(\gamma_0\gamma_5 \gamma^\nu
   \gamma_5 \sigma^{12})+\cos(eBs)\sinh(eEs)
   \tr(\gamma_0 \gamma_5 \gamma^\nu
   \gamma_{5}\sigma^{12})\Bigr]\,.
\end{align}
Keeping the relevant terms in the spin factor, and closing the contours in
$\Gamma^<$ and $\Gamma^>$, we find for Eq.~\eqref{eq:j5_in_trace},
\begin{equation}
  j^\mu_5=\lim_{x\rightarrow y} i
  \frac{e^2EB}{4\pi^2}  \partial^3 \theta(z_3) 
  I_h(z) \delta^\mu_3 \,.
\end{equation}
Let us pause the above calculation to digress on the emergence 
of real-time. This in fact stems from the $\theta(\pm z_3)$ terms
in $\int_{in}$. When acted upon by the partial derivative the 
resulting delta function is a measure of the phase space in our 
system and real-time, $t$,
dependence~\cite{Nikishov:1969tt,fradkin1991quantum}:
\begin{equation}
   \lim\limits_{y\rightarrow x} \delta(z_3)=\lim\limits_{y\rightarrow x}
   \int \frac{dp_{3}}{2\pi}e^{ip^3 z^3}=\frac{eEt}{2\pi}.
   \label{eq:delta_time}
\end{equation}

Such an identification is just; we can approach this through both 
an examination of the canonical and kinetic momenta 
differences~\cite{TANJI20091691}, and also through a look at the Dirac 
equation~\cite{Nikishov_long}. For the former, let us consider a wavepacket
 perspective. A magnetic
field and hence electric field will give harmonic oscillator solutions to the
Dirac equation. For the electric fields, the wavepacket would have its
solutions in time shifted by the canonical momentum, $p_3/eE$, 
and also energies would be independent of $p_3$. 
Then assuming for some initial time the kinetic momentum of the produced 
pairs would be zero would imply $p_3=eEt$, which gives $t$ as a total time 
measure of the system.
One could then anticipate an integration over all canonical momentum
as being analogous to one over all time. However in this argument, and
also as it pertains to Eq.~\eqref{eq:delta_time}, one must emphasize
that a picture of particle production is valid at any time. We can see this
in the fact that solutions of the Dirac equation, are in fact valid at any time,
not just in the out (or in) asymptotic states~\cite{TANJI20091691}. 
Therefore, Eq.~\eqref{eq:delta_time} is only valid for operators
evaluated at an asymptotic time, i.e., $a^{out,\,in}_{n},b^{out,\,in}_{n}$.

In an asymptotic state expansion of the Dirac equation in an electric
field--or with parallel magnetic field--one encounters a parabolic
cylinder function with time dependence~\cite{Nikishov_long}, e.g.,
$D_{a-1}[\sqrt{eE}^{-1}(1-i)(p_3-eE x_0)]$, for complex coefficient
$a$. 
The complex parabolic cylinder functions have distinct particle and 
anti-particle pictures at asymptotic times. Let's take for example one with a 
particle identification at $x_0\rightarrow \infty$. Taking the same solution 
but at $x_0 \rightarrow - \infty$, one can see an admixture of particle
and antiparticle states. Therefore, one can indicate a time 
intervals in which the particle and antiparticle states are fully 
defined~\cite{Nikishov_long}. This time interval is 
centered about $x_0=p_3/eE$ in 
the above. revealing when pair production occurs. 
See Ref.~\cite{Nikishov_long} for further details.

Finally, using Eq.~\eqref{eq:delta_time}
and carefully evaluating the Heaviside function arguments using notations
given in Sec.~\ref{sec:introduction} one can find for the 
in-in real-time chiral density from the Schwinger mechanism 
as\cite{Warringa:2012bq,Copinger:2018ftr},
\begin{equation}
  j^\mu_5=\frac{e^2EB\, t}{2\pi^2}
  \exp\bigl(-\frac{\pi m^2}{eE}\bigr)\delta^\mu_3 \,,
\end{equation}
in agreement with the pseudoscalar and axial Ward identity calculations given 
in Eqs.~\eqref{eq:Pin} and ~\eqref{eq:awi_ii} respectively.
Let us also point out that the chirality production via the Schwinger
mechanism has been extended to dynamically assisted configurations
enhancing the rate of 
chirality production~\cite{PhysRevResearch.2.023257}.

\section{Chiral Magnetic Effect Current}
\label{sec:cme}

We have established the role the Schwinger mechanism plays on the chiral 
anomaly through the axial Ward identity above, and it would be instructive
to examine the CME as well. As expected similar 
behavior exists for the CME and chiral vector
current in homogeneous fields. As advertised in Sec.~\ref{sec:introduction} 
and as argued in Ref.~\citen{Fukushima:2010vw}, 
there is no need for artificial placement of 
a chiral chemical potential to see the CME. Let us demonstrate that here 
and in so doing confirm in and out-of equilibrium characteristics of the CME.

The vector currents in in-out and in-in formalism may be cast in proper time
as before,
\begin{align}
   \bar{j}^\mu&:=\langle \bar{\psi}\gamma^\mu \psi\rangle 
   = i \lim_{y\to x} \tr\bigl[\gamma^\mu \sprop(x,y)\bigr] 
   = -i\lim_{y\to x}\tr[\gamma^\mu 
   \,\slashed{D}_x \int_0^\infty ds\, g(x,y,s)]\,, \label{eq:cme_in} \\
    j^\mu&:=\llangle \bar{\psi}\gamma^\mu\psi\rrangle
    = i \lim_{y\to x} \tr\bigl[\gamma^\mu \Sprop(x,y)\bigr]
    = -i\lim_{y\to x}\tr[\gamma^\mu 
   \,\slashed{D}_x \int_{in} ds\, g(x,y,s)]\,.
    \label{eq:cme_out}
\end{align}
Using similar steps as outlined for Eq.~\eqref{eq:res_D}, 
(where we found 
that the covariant derivative acting on the kernel in the $x\rightarrow y$
limit vanished due to a translational invariance), one can find that the
Euclidean equilibrium CME current, $\bar{j}^3$ vanishes,
as is understood in condensed matter 
applications~\cite{Yamamoto:2015fxa}. Then as before, one can
see the emergence of the CME in an out-of equilibrium context, 
here sourced through the Schwinger mechanism.

Calculations for the real-time CME follow closely to those done 
for the chiral density, Eq.~\eqref{eq:j5_in_trace}, 
therefore let us simply outline 
some key steps. As before, we can eliminate terms without poles coming
from the spin and boson path integral factors. The Dirac trace here is
\begin{align}
   &\tr\gamma^{\mu}\slashed{D}
   e^{-\frac{i}{2}eF\sigma s}
   =4D_{\nu}\Bigl\{\cos(eBs)\cosh(eEs)
   g^{\mu\nu} \nonumber \\
   &-\sin(eBs)\cosh(eEs)(-g^{\mu1}g^{\nu2}
   +g^{\mu2}g^{\nu1})-\cos(eBs)\sinh(eEs)\epsilon^{\mu\nu12}\Bigr\}\,.
\end{align}
Let us also note that $z_0$ dependence vanishes in the $x\rightarrow y$ 
limit since $\partial^0 \theta(z_3^2-z_0^2)=0$. We essentially find 
Eq.~\eqref{eq:j5_in_trace}, however, with a sum over the Landau 
levels corresponding to the $\coth(\pi B/E)$ term included:
\begin{align}
   j^\mu   &=\frac{e^2EB\,t}{2\pi^2}\coth\bigl(\frac{\pi B}{E}\bigr)
   \exp\bigl(-\frac{m^2\pi}{eE}\bigr)\,\delta_3^\mu 
   =2\omega t\,\delta_3^\mu\,.
   \label{eq:cme_final}
\end{align}
A current emerges in accordance with
Schwinger's formula given in Eq.~\eqref{eq:schwinger_formula} under 
the LLLA of the above.
We find while the CME vanishes in Euclidean equilibrium, it reemerges
out-of equilibrium in a real-time picture 
through the Schwinger mechanism in QED.
The above expression and connection to the CME through the Schwinger
mechanism was first examined in Ref.~\citen{Fukushima:2010vw}, 
relying on a Lorentz
transformation of Schwinger's formula. And that a current is generated from 
the Schwinger mechanism is indeed well 
known~\cite{fradkin1991quantum,PhysRevD.78.045017,TANJI20091691}.
While all Landau  levels
 have been kept in the above analysis, let us emphasize
that the CME should only appear as a result of generated chirality. However
we saw in Eqs.~\eqref{eq:Pin} and~\eqref{eq:awi_ii} that only the LLLA 
contributed to a net chirality. 

We also noticed that these currents can also be computed through the equal-
time Wigner function approaches\cite{Zhuang:1995jb,Zhuang:1998bqx, 
Hebenstreit:2010vz,Mao:2018jdo, Sheng:2018jwf, Sheng:2019ujr}.
Having seen the importance the Schwinger mechanism
plays on both the anomaly and the CME, let us examine its effects on the chiral 
condensate.

\section{Chiral Condensate}
\label{sec:condensate}

The chiral condensate possesses an interesting interplay with the chiral
symmetry, and besides its finiteness giving rise to a baryon mass, the chiral
condensate may be enhanced in an external magnetic field in what is known
as magnetic catalysis~\cite{Klimenko:1992ch,Gusynin:1994re,
Gusynin:1994xp,Shovkovy:2012zn}. Then it is an interesting extension,
we explore here, to evaluate the chiral condensate in a strong electric
field such that Schwinger pair production be producible. Also, to what effect 
does the out-of equilibrium process entail for dynamical mass; this too we
can address here. 

The chiral condensates both in and out-of equilibrium respectively are
\begin{align}
    \bar{\Sigma}&\coloneqq \langle\bar{\psi}\psi\rangle=
    i\lim_{y\to x} \tr\bigl[ \sprop(x,y)\bigr]\,, \\
    \Sigma&\coloneqq \llangle\bar{\psi}\psi\rrangle=
    i\lim_{y\to x} \tr\bigl[ \Sprop(x,y)\bigr]\,.
    \label{eq:sigma_ii}
\end{align}
We first treat the magnetic catalysis case; this is simply the one 
with $E=0$, and hence either of the expression above may be used.
Let us also employ the LLLA as was used for the chiral density fluctuations.
Then the chiral condensate can be found as
\begin{align}
  \bar{\Sigma}\bigr|_{E=0} &
  = -\frac{eB}{4\pi^2}\, m\int_0^\infty
  \!\frac{ds}{s} e^{-i m^2 s} \cot(eB s) \nonumber\\
   &= -\frac{eB}{4\pi^2}\, m\int_{m^2/\Lambda^2}^\infty
  \!\frac{ds}{s} e^{-m^2 s} \coth(eB s)  \nonumber\\
  & \simeq -\frac{eB}{4\pi^2}\, m\, \Gamma[0,m^2/\Lambda^2]\,.
  \label{eq:catalysis}
\end{align}
An ultraviolet cutoff has been introduced in the second step, where also
a rotation in $s\rightarrow -is$ has been done--making a connection to more 
familiar constructions~\cite{Gusynin:1994re,Gusynin:1994xp}.
Magnetic catalysis emerges for small $m$ from an infinite negative 
curvature after solving the gap equation for the condensate stemming 
from the logarithmic singularity in the condensate, 
$\Gamma[0,m^2/\Lambda^2]\simeq 
   -\gamma_{\rm E}+\ln(\Lambda^2/m^2)$,
where $\gamma_{\rm E}$ is the Euler-Mascheroni constant. 
Let us now examine how the condensate behaves under a parallel
electric field starting with the Euclidean equilibrium case first.

The in-out chiral condensate can be found straightforwardly:
\begin{align}
  \bar{\Sigma} &
  = -\frac{e^2EB}{4\pi^2}\, m\int_0^\infty
   ds\, e^{-i m^2 s} \cot(eB s) \coth(eEs)\nonumber \\
   &= -\frac{e^2EB}{4\pi^2}\, m\int_{m^2/\Lambda^2}^\infty
   ds\, e^{-m^2 s} \coth(eB s)\cot(eEs)\nonumber  \\
   &\simeq -\frac{e^2EB}{4\pi^2}\, m\int_{m^2/\Lambda^2}^\infty
  ds\, e^{-m^2 s} \cot(eEs) \nonumber \\
  &\simeq -\frac{eB}{4\pi^2}\, 
  m \biggl[\ln\frac{\Lambda^2\, e^{-\gamma_{\rm E}}}{2eE}
    - \mathrm{Re}\psi\Bigl(\frac{i m^2}{2eE}\Bigr)
    - \frac{i\pi}{e^{\pi m^2/(eE)}-1}\biggr]\,.
\label{eq:condensate_in}
\end{align}
In addition to the LLLA, we also approximate for large $\Lambda$,
$e^{-m^2/\Lambda^2}\sim 1$ and also thereafter only leading
order contributions of $\Lambda^2$ have been kept. $\psi(x)$
here is the digamma function. We can see in 
Eq.~\eqref{eq:condensate_in} that the logarithmic singularity 
with respect to $m^2$ has disappeared~\cite{Wang:2018gmj}. 
Furthermore, there is a suppression of the condensate with the
inclusion of the electric field; this we will explore in greater
depth with the realization of Schwinger pair production provided
by the in-in construction. Also in Eq.~\eqref{eq:condensate_in}
we see there is an imaginary piece--as alluded to in
Sec.~\ref{sec:expectation}. What is interesting is the form 
of the imaginary part resembles a bosonic-like distribution,
with ``temperature" in proper time given by $\pi /(eE)$.
A similar distribution is also in fact present for the squared
matrix element predicting the probability for a single particle
pair to be found due to the Schwinger mechanism. That a 
``temperature" arises highlights a non-equilibrium nature,
that we examine in the real-time picture below. 
Let us also mention that a temperature arises in the worldline
picture from a dynamical gauge field in addition to the background
gauge field through sphaleron 
transitions~\cite{PhysRevD.96.076002,PhysRevD.98.056022}.
Furthermore,
the real-time quantity is real as expected. Let us also
point out that complex features have also been seen in QFTs
under a finite $\theta$~\cite{Boer:2008ct,Mameda:2014cxa}.  
And, a topological $\theta$ and our fields, Eq.~\eqref{eq:homogeneous}
share similar quantum numbers, therefore complex observables
would be anticipated. Last, let us also mention that one can recover 
Eq.~\eqref{eq:catalysis} from Eq.~\eqref{eq:condensate_in}
by noting the asymptotic expansion, 
$\psi(x)\sim \ln x - 1/2x$ for large $x$. 

\begin{figure}
\centering
  \includegraphics[width=0.8\columnwidth]{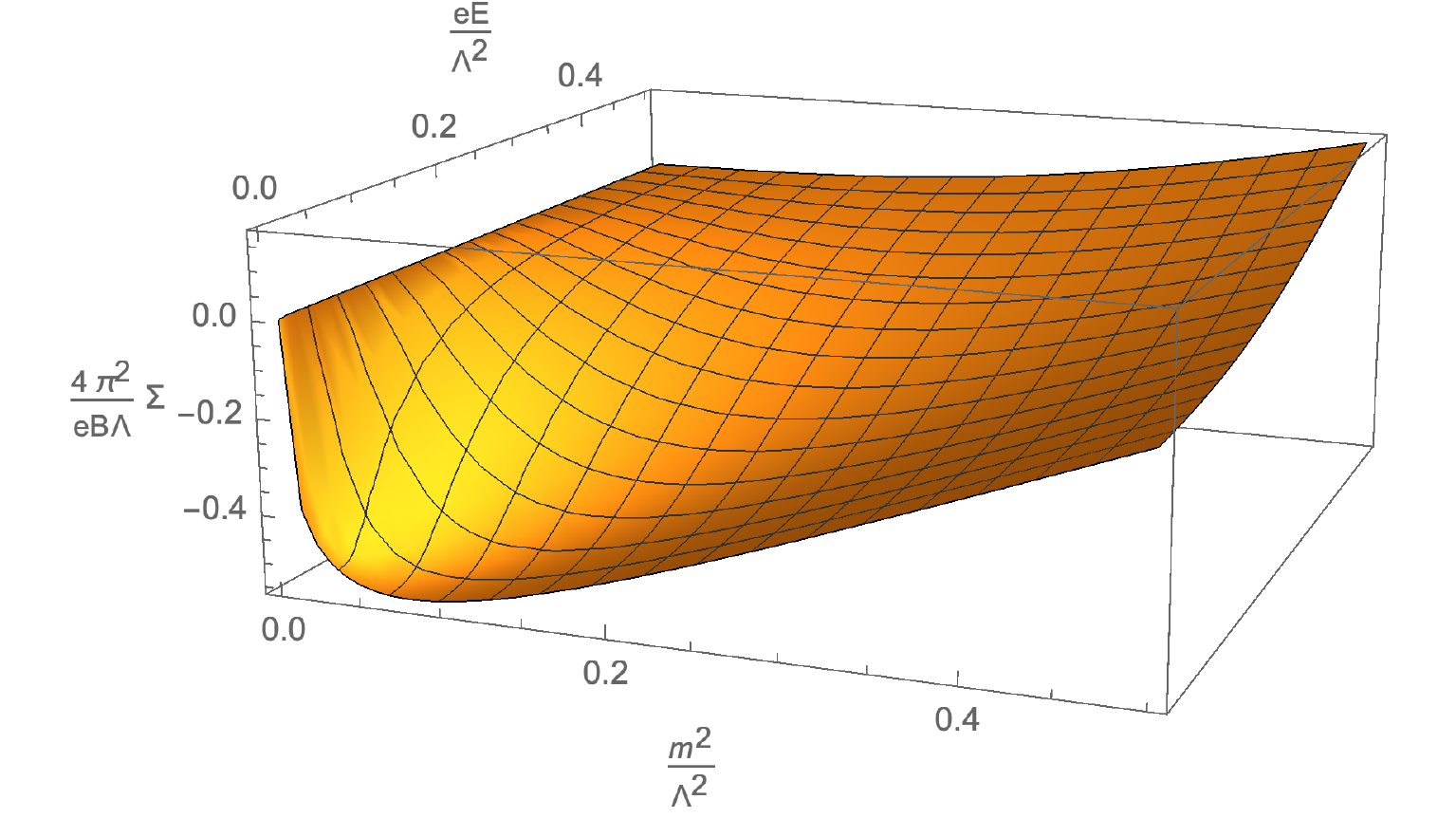}
  \caption{In-in (out-of equilibrium) chiral condensate, 
  Eq.~\eqref{eq:condensate_out} in parallel electric and magnetic fields. The
  dimensionless condensate is divided by $eB/4\pi^2$ and the scale 
  $\Lambda$. Condensate is depicted for background electric fields, 
  $eE/\Lambda^2$, and mass, $m^2/\Lambda^2$. For large $E$,
  observe a melting of the condensate, $\Sigma \rightarrow 0$, and
  restoration of the chiral symmetry. The constituent mass is decreased
  from the Schwinger mechanism.}
  \label{fig:melting}
\end{figure}

We find for the in-in chiral condensate, Eq.~\eqref{eq:sigma_ii},
\begin{align}
  \Sigma &= -\frac{e^2EB}{4\pi^2}\, m\int_{in}
   ds\, e^{-i m^2 s} \cot(eB s) \coth(eEs) \nonumber \\
  &= -\frac{e^2 EB}{4\pi^2}\, m\int_{1/\Lambda^2}^{\pi/eE-1/\Lambda^2}
  \!\!\!\!\!\!\!\! ds\, e^{-m^2 s} \coth(eBs)\cot(eEs) \nonumber\\
  &\simeq \Bigl[ 1 - e^{-\pi m^2/(eE)} \Bigr] \,\text{Re}\,
  \bar{\Sigma}\,,
  \label{eq:condensate_out}
\end{align}
under the LLLA~\cite{Copinger:2018ftr}. 
As with the equilibrium case, Eq.~\eqref{eq:condensate_in},
we see there are divergences; there are two poles here at $s=0, -i\pi/eE$. 
However both UV divergences are approached the same way as for
Eq.~\eqref{eq:condensate_in}, therefore we regulate them similarly.
The out-of equilibrium chiral condensate is depicted in Fig.~\ref{fig:melting}.
In contrast to magnetic catalysis, 
we see that with the addition of an
electric field the chiral condensate is 
weakened~\cite{SUGANUMA1991470,10.1143/ptp/90.2.379,
CAO2020135477}, 
which acts as an 
inverse magnetic catalysis effect. 
One may
understand this process intuitively: While a magnetic field would act to 
strengthen the condensate through spin alignment, an electric field 
would act to pull the condensate apart, in effect weakening it.

A melting of the chiral condensate in an electric field might 
be observable in a condensed matter system. In contrast to QED,
the energy gap in a Weyl semimetal is small, and it has been reasoned 
the Schwinger mechanism may be measured there~\cite{Vajna:2015qra}.
Also, magnetic catalysis might be visible in Weyl
semimetals~\cite{PhysRevB.92.125141}, and thus the semimetal
may prove a vital means of accessing the melting behavior.
One can also see Ref. \citen{Cao_2020, Cao_2020_PRD} for recent 
discussions in the electric field dependent chiral condensation.

\section{Conclusions}
\label{sec:conclusions}

Chirality generation from the QFT vacuum via the Schwinger mechanism 
has been examined. We have demonstrated the importance of vacuum 
states for the determination of expectation values. Notably, we showed
that in-out expectation valued observables coincide with a scenario of
Euclidean equilibrium, and in-in expectation valued observables
predict a situation out-of equilibrium. With an understanding of the
difference of vacuum states for the production of chirality through
Schwinger pair production, it was demonstrated how a heuristic 
picture of the process is indeed accurate. And also, with the
understanding, it is reasoned the anomaly and related quantities
 should vanish in equilibrium.

It was found the pseudoscalar condensate, and hence axial Ward identity,
by virtue of the Schwinger mechanism acquired mass dependence: A
characteristic exponentially quadratic mass suppression was 
calculated out-of equilibrium, as too for a CME current.
Also for the CME current as well as the chiral density current, a 
real-time (as would be expected from the in-in
formalism) dependence emerged from a phase space factor. For 
the chiral density current, this was in accordance with the axial Ward 
identity. 
A chiral condensate with Schwinger mechanism effects was also discussed,
where it was shown the condensate weakens, even vanishing, in
a background electric field, in effect, acting as an inverse magnetic catalysis.

Here the beginnings of chirality generation by the Schwinger mechanism
have been outlined, however future work is necessary to both expand 
and deepen our understanding. A notable shortcoming in the 
analysis presented
here is the usage of homogeneous Abelian fields.  While a general framework
exists for handling in-out expectation values in a worldline picture, it is 
important to extend
the worldline formalism for in-in expectation values to arbitrary fields. It 
would then
be of interest to rigorously confirm the dependence of the Schwinger
mechanism on the chiral anomaly in a non-Abelian background with
 topological winding number. And also for the cancellation
of the anomaly in Euclidean equilibrium, it would be important to analyze 
similar non-trivial field types.

\section*{Acknowledgments}
The authors would like to thank Gaoqing Cao, Kenji Fukushima, Xu-Guang 
Huang, and Hidetoshi Taya for helpful discussions and comments.

\bibliography{references}

\end{document}